\documentclass[twoside,twocolumn,9pt]{article}
\usepackage{extsizes}
\usepackage[super,sort&compress,comma]{natbib}
\usepackage[version=3]{mhchem}
\usepackage[left=1.5cm, right=1.5cm, top=1.785cm, bottom=2.0cm]{geometry}
\usepackage{balance}
\usepackage{widetext}
\usepackage{times,mathptmx}
\usepackage{sectsty}
\usepackage{graphicx}
\usepackage{lastpage}
\usepackage[format=plain,justification=raggedright,singlelinecheck=false,font={stretch=1.125,small,sf},labelfont=bf,labelsep=space]{caption}
\usepackage{float}
\usepackage{fancyhdr}
\usepackage{fnpos}
\usepackage[english]{babel}
\usepackage{array}
\usepackage{droidsans}
\usepackage{charter}
\usepackage[T1]{fontenc}
\usepackage[usenames,dvipsnames]{xcolor}
\usepackage{setspace}
\usepackage[compact]{titlesec}

\definecolor{cream}{RGB}{222,217,201}

\usepackage{bm,xstring,xfrac}   %
\usepackage{hyperref}
\usepackage{amsmath,amssymb}
\newcommand{\ie}{i.e.,~}
\newcommand{\eg}{e.g.,~}
\newcommand{\secref}{Section~\ref}
\newcommand{\etal}{\textit{et al.} }

\newcommand{\Tr}[1] {\textrm{Tr}\left[#1\right]}

\newcommand{\tsr}[1] {\mathbf{#1}}
\newcommand{\vtr}[1] {\mathbf{#1}}

\newcommand{\figref}{Fig.~\ref}
\newcommand{\figsref}{Figs.~\ref}
\renewcommand{\eqref}{Eq.~\ref}
\newcommand{\eqsref}{Eqs.~\ref}

\newcommand{\GQ}[0]{G_{\rm Q}}
\newcommand{\la}[0]{\ell_\alpha}
\newcommand{\lQ}[0]{\ell_Q}

\newcommand{\lqq}[0]{\ell_\theta}
\newcommand{\ld}[0]{\ell_d}
\newcommand{\lv}[0]{\ell_v}
\newcommand{\lw}[0]{\ell_\Omega}

\begin{document}

\pagestyle{fancy}
\thispagestyle{plain}
\fancypagestyle{plain}{

\fancyhead[C]{\includegraphics[width=18.5cm]{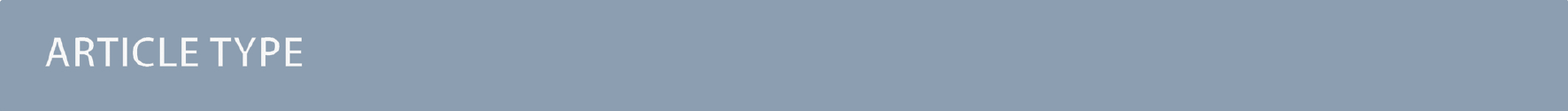}}
\fancyhead[L]{\hspace{0cm}\vspace{1.5cm}\includegraphics[height=30pt]{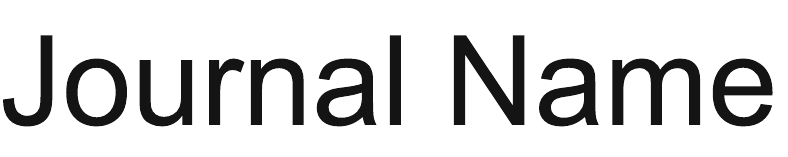}}
\fancyhead[R]{\hspace{0cm}\vspace{1.7cm}\includegraphics[height=55pt]{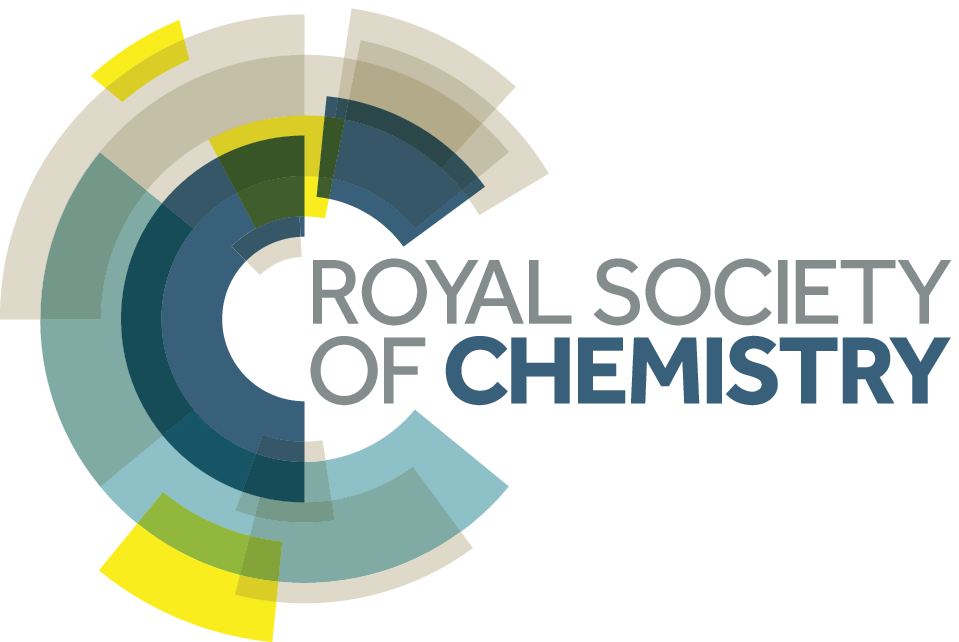}}
\renewcommand{\headrulewidth}{0pt}
}
\makeFNbottom
\makeatletter
\renewcommand\LARGE{\@setfontsize\LARGE{15pt}{17}}
\renewcommand\Large{\@setfontsize\Large{12pt}{14}}
\renewcommand\large{\@setfontsize\large{10pt}{12}}
\renewcommand\footnotesize{\@setfontsize\footnotesize{7pt}{10}}
\makeatother

\renewcommand{\thefootnote}{\fnsymbol{footnote}}
\renewcommand\footnoterule{\vspace*{1pt}%
\color{cream}\hrule width 3.5in height 0.4pt \color{black}\vspace*{5pt}}
\setcounter{secnumdepth}{5}

\makeatletter
\renewcommand\@biblabel[1]{#1}
\renewcommand\@makefntext[1]%
{\noindent\makebox[0pt][r]{\@thefnmark\,}#1}
\makeatother
\renewcommand{\figurename}{\small{Fig.}~}
\sectionfont{\sffamily\Large}
\subsectionfont{\normalsize}
\subsubsectionfont{\bf}
\setstretch{1.125} %
\setlength{\skip\footins}{0.8cm}
\setlength{\footnotesep}{0.25cm}
\setlength{\jot}{10pt}
\titlespacing*{\section}{0pt}{4pt}{4pt}
\titlespacing*{\subsection}{0pt}{15pt}{1pt}
\fancyfoot{}
\fancyfoot[LO,RE]{\vspace{-7.1pt}\includegraphics[height=9pt]{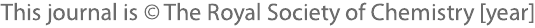}}
\fancyfoot[CO]{\vspace{-7.1pt}\hspace{13.2cm}\includegraphics{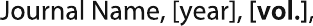}}
\fancyfoot[CE]{\vspace{-7.2pt}\hspace{-14.2cm}\includegraphics{head_foot/RF}}
\fancyfoot[RO]{\footnotesize{\sffamily{1--\pageref{LastPage} ~\textbar  \hspace{2pt}\thepage}}}
\fancyfoot[LE]{\footnotesize{\sffamily{\thepage~\textbar\hspace{3.45cm} 1--\pageref{LastPage}}}}
\fancyhead{}
\renewcommand{\headrulewidth}{0pt}
\renewcommand{\footrulewidth}{0pt}
\setlength{\arrayrulewidth}{1pt}
\setlength{\columnsep}{6.5mm}
\setlength\bibsep{1pt}
\makeatletter
\newlength{\figrulesep}
\setlength{\figrulesep}{0.5\textfloatsep}

\newcommand{\topfigrule}{\vspace*{-1pt}%
\noindent{\color{cream}\rule[-\figrulesep]{\columnwidth}{1.5pt}} }

\newcommand{\botfigrule}{\vspace*{-2pt}%
\noindent{\color{cream}\rule[\figrulesep]{\columnwidth}{1.5pt}} }

\newcommand{\dblfigrule}{\vspace*{-1pt}%
\noindent{\color{cream}\rule[-\figrulesep]{\textwidth}{1.5pt}} }

\makeatother
\twocolumn[
  \begin{@twocolumnfalse}
\vspace{3cm}
\sffamily
\begin{tabular}{m{4.5cm} p{13.5cm} }

  \includegraphics{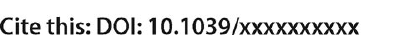} & \noindent\LARGE{\textbf{Correlation lengths in hydrodynamic models of active nematics$^\dag$}} \\
\vspace{0.3cm} & \vspace{0.3cm} \\

    & \noindent\large{\textbf{Ewan J. Hemingway,\textit{$^{a\ast\ddag}$} Prashant Mishra,\textit{$^{b\ddag}$} M. Cristina Marchetti,\textit{$^{b}$} and Suzanne M. Fielding\textit{$^{a}$}}}\\

\includegraphics{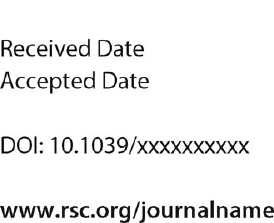} & \noindent\normalsize{We examine the scaling with activity of the emergent length scales that control the nonequilibrium dynamics of an active nematic liquid crystal, using two popular hydrodynamic models that have been employed in previous studies. In both models we find that the chaotic spatio-temporal dynamics in the regime of fully developed active turbulence is controlled by a single active scale determined by the balance of active and elastic stresses, regardless of whether the active stress is extensile or contractile in nature. The observed scaling of the kinetic energy and enstropy with activity is consistent with our single-length scale argument and simple dimensional analysis. Our results provide a unified understanding of apparent discrepancies in the previous literature and demonstrate that the essential physics is robust to the choice of model. } \\

\end{tabular}

 \end{@twocolumnfalse} \vspace{0.6cm}

  ]
\renewcommand*\rmdefault{bch}\normalfont\upshape
\rmfamily
\section*{}
\vspace{-1cm}

\footnotetext{\textit{$^{a}$~Department of Physics, Durham University, Science Laboratories, South Road, Durham, DH1 3LE, UK}}
\footnotetext{\textit{$^{b}$~Physics Department and Syracuse Soft Matter Program, Syracuse University, Syracuse, NY 13244, USA}}

\footnotetext{\dag~Electronic Supplementary Information (ESI) available: videos of the two regimes of activity driven turbulence in Model II. See DOI: \href{http://dx.doi.org/10.1039/C6SM00812G}{10.1039/C6SM00812G}.}

 \footnotetext{\ddag~ These authors contributed equally to this work.}

\section{Introduction}

Active systems are continuously driven out of equilibrium by energy injected at the local scale, resulting in collective motion at the large scale.  Examples include bacterial colonies, \textit{in vitro} extracts of cytoskeletal filaments and associated motor proteins, and monolayers of vibrated granular matter~\cite{Marchetti:2013}.  Much interest has recently focused on active nematics - systems composed of active units with head-tail symmetry that can order into states with nematic liquid crystalline order, with the activity giving rise to a rich variety of collective phenomena. These include spontaneous laminar flow~\cite{Voituriez:2005,Marenduzzo:2007,Giomi:2008}, large density fluctuations~\cite{Ramaswamy:2003,Mishra:2006,Narayan:2007}, pattern formation~\cite{Giomi:2011,Giomi:2012}, spontaneous unbinding of topological defects~\cite{Pismen:2013,Giomi:2013,Shi:2013,luca-RS}, and active turbulence~\cite{Thampi:2013, Thampi:2014a,thampi-RS, Giomi:2014,Thampi:2015}.

Theoretical interest in active nematics has been fueled by the observation of spontaneously flowing and turbulent states in suspensions of microtubule-kinesin bundles confined at an oil-water interface~\cite{Sanchez:2012, Keber:2014, Henkin:2014}. Other recent experimental realizations  were obtained by immersing living swimming bacteria (specifically, \textit{E.~coli}) in a lyotropic liquid crystal~\cite{Zhou:2014} and by plating dense layers of fibroblasts on substrates~\cite{Duclos:2014}. Microtubule bundles and \textit{E.~coli} exert active extensile force dipoles on their surroundings, while the fibroblasts exert contractile force dipoles. In both systems active stresses couple to orientational order and induce flows and defect unbinding, with qualitatively similar non-equilibrium dynamics, albeit on very different time scales.

All these experiments have indicated that the appearance of spatio-temporal chaos in active nematics is accompanied by the proliferation of topological defects, which may mediate the onset of turbulence even at negligible values of the Reynolds number~\cite{Giomi:2013,Thampi:2013}.  In active fluids, distortions of the local orientational order produce local shear flows that enhance the orientational deformation, ultimately leading to the unbinding of defect pairs. In nematics, these consist of pairs of $\pm 1/2$ disclinations - specific distortions of the orientational order that are the signature of the underlying broken symmetry.  It was additionally shown~\cite{Giomi:2013} that the dynamics of topological defects in active systems depends on the nature of the forcing, \ie whether the force dipole is contractile or extensile. This result provides a useful criterion for inferring the nature of active forces in systems with nematic symmetry~\cite{Giomi:2013,thampi-RS,Gao:2015}.

In spite of much theoretical work, discrepancies still exist in the literature over the nature of the characteristic length scales that control the spontaneous proliferation and annihilation of topological defects, and the resulting dynamics in the so-called turbulent state. In particular, the dependence of such length scales on the strength of the active forcing, $|\alpha|$,  remains unclear.  A recent numerical study by Giomi\cite{Giomi:2014} examined the statistics of the activity-driven turbulent phase in two-dimensional nematic films by measuring the distribution of vortex sizes for a selection of activities. This work provided evidence that the key physics is determined by a single active length scale, $\la$, proportional to $|\alpha|^{-1/2}$. In contrast, in a closely related work, but on a different continuum model of a quasi-$2D$ nematic, Thampi \etal performed a detailed study that measured several orientational and hydrodynamical correlation lengths, suggesting that the length scale of structure in the fluid instead scales as $|\alpha|^{-1/4}$.

An important aim of this work is to provide a unified understanding of these previous, apparently conflicting reports. In order to do so, we consider two different but related models of active nematic liquid-crystal hydrodynamics that are commonly used in the literature.  The dynamics of these models is compared using two independently developed numerical codes. By varying the key dimensionless parameters over several decades, we obtain data to support the conjecture that, in both models, the mean defect spacing in the regime of full developed active turbulence is set by the length scale $\ell_\alpha\sim \left(K/|\alpha|\right)^{1/2}$, defined by the balance of active and elastic stresses. Here $|\alpha|$ is the magnitude of the active stress (commonly referred to as the activity) and $K$ parametrizes the free energy penalty that results from spatial variations in the director field~\cite{Marenduzzo:2007,Giomi:2008, Marchetti:2013, luca-RS, Giomi:2014,DeGennes:1993}. We show that this result holds for both extensile and contractile systems, in both the flow aligning and the flow tumbling regimes. This active length scale also controls the onset of spontaneous laminar flow in an active film, a phenomenon that has been referred to in the literature as the spontaneous flow instability~\cite{Voituriez:2005}. Our study  provides the first explicit demonstration that distinct constitutive models produce the same emergent length scale, \ie they both produce quantitatively consistent scaling relations. We also demonstrate a regime of less highly developed turbulence in which a weaker scaling $\la \sim |\alpha|^{-1/4}$ appears consistent with our numerical data.

In many experimental realizations active nematics are confined to quasi two-dimensional geometries, \eg on the surface of lipid vesicles~\cite{Keber:2014}, in flattened water-in-oil droplets~\cite{Sanchez:2012}, in thin-films \cite{Sokolov2007}, or squeezed between parallel glass plates~\cite{Zhou:2014}. It is worth noting that the presence of confining walls (with no-slip boundary conditions) in the last of these modifies hydrodynamic interactions between active particles \cite{Wensink2012}; of the above experiments, free-standing films may then provide the closest experimental realisation of our system. Clearly, in any numerical study, it is important to define carefully the considered dimensionality. In what follows we denote by $D$ the number of dimensions in which the relevant fields (nematic order parameter tensor, fluid velocity, {\it etc.}) are allowed to vary; and separately by $d$ the number of dimensions in which the nematic director is allowed to develop non-zero components.  We shall perform two different studies. In the first we take a strictly two-dimensional model of an active nematic sheet, in which the order parameter tensor $\tsr{Q}$ is allowed to develop non-zero components only in the $x-y$ directions ($d=2$); and physical quantities are likewise allowed to vary only in the $x-y$ plane ($D=2$). In the second study we consider a three-dimensional nematic ($d=3$) but in which all quantities are nonetheless still assumed to be spatially homogeneous in the direction of the layer thickness ($D=2$).  In the latter case the director can in principle point out of the simulated plane, and indeed this effect has been reported in a previous numerical study of active nematics in a cylindrical capillary \cite{Ravnik2013}. In that work confinement in the plane of simulation plays an important role whereas our study deliberately focuses on the bulk dynamics where we do not observe out-of-plane motion. Finally it we note that accurately resolving turbulent active hydrodynamics is computationally demanding, especially when performing large sweeps of parameter space. This necessarily restricts us to a 2D study, as is the case with many other studies of active turbulence \cite{Marenduzzo:2007,Thampi:2013,Giomi:2014}.

The paper is structured as follows. In \secref{sec:models} we define the equations of motion for both models, and outline the parameter ranges that we explore for each. In \secref{sec:char_lengths}, we define the observable length scales that can be used to characterize the fluid structure, and discuss the physical reasoning behind their definition.  The results of our study are presented in \secref{sec:results}. In \secref{sec:discussion} we provide a comparison with other works and offer our conclusions.

\section{Models}
\label{sec:models}

In $D=2$ spatial dimensions we consider an incompressible uniaxial active nematic liquid crystal with a director that can orient in $d$ dimensions, with $d=2,3$ in our two respective studies. The nematic orientational order is parametrized by a symmetric and traceless tensor field $Q_{ij} = \frac{Sd}{2}(n_i n_j - \frac{\delta_{ij}}{d})$, where $S$ is the order parameter magnitude and the director $\vtr{n}$ is a headless unit vector that characterizes the direction of broken orientational symmetry. The nematic is embedded in an incompressible fluid of constant density, $\rho$, and constant viscosity, $\eta$.  The fluid velocity field is denoted by $\vtr{v}$. The associated pressure field $p$ is determined by the incompressibility condition $\bm\nabla\cdot\vtr{v} = 0$.

The equations of motion for an active nematic are derived from the well-known hydrodynamic equations for a passive liquid-crystal~\cite{DeGennes:1993}
\begin{align}
  \rho D_t\mathbf{v} &= \eta \nabla^2 \mathbf{v} - \bm\nabla p +\bm\nabla\cdot\bm\Sigma^T\;,\label{eq:navier_stokes}\\
  D_t \mathbf{Q} &= 2[\mathbf{Q}\cdot \bm\Omega]^A+\mathbf{M}^{(d)}(\mathbf{D},\mathbf{Q})+\frac{1}{\gamma}\mathbf{H}\;,
  \label{eq:nematic dynamics}
\end{align}
where $D_t= \left(\partial_t + \mathbf{v}\cdot\bm\nabla\right)$ is the material derivative and $\gamma$ is a rotational viscosity.  Here $\mathbf{D}$ and $\bm\Omega$ denote the symmetric and antisymmetric parts of the rate of strain tensor $(\nabla \rm{v})_{ij} \equiv \partial_i \rm{v}_j$, respectively, with $D_{ij}=\frac12\left(\partial_i\rm{v}_j+\partial_j\rm{v}_i\right)$ and $\Omega_{ij}=\frac12\left(\partial_i\rm{v}_j-\partial_j\rm{v}_i\right)$. For other tensors the transpose, symmetric, antisymmetric and traceless parts are denoted by the superscripts $\dag$, $S$, $A$ and $T$, respectively. For example, $[\mathbf{B}]^A =\frac12\left[ \mathbf{B} - \mathbf{B}^\dag\right]$.

The relaxation dynamics of the alignment tensor in \eqref{eq:nematic dynamics} is governed by the molecular field, $\mathbf{H}=-[\frac{\delta F}{\delta \mathbf{Q}}]^{ST}$, in which the Landau-de Gennes free energy~\cite{DeGennes:1993}, $F=\int dV (f_b+f_d)$, is the sum of contributions from a bulk free energy density
\begin{equation}
  \begin{split}
  f_b = \GQ\left\{\frac{A}{2}\Tr{\mathbf{Q}^2}+\frac{B}{3}\Tr{\mathbf{Q}^3}+\frac{C}{4}\Tr{\mathbf{Q}^2}^2\right\}\;,
  \end{split}
  \label{eq:landau}
\end{equation}
and the distortion free energy density
\begin{equation}
  f_d = \frac{K}{2}\partial_i Q_{jk}\partial_i Q_{jk}\;.
  \label{eq:distortion}
\end{equation}
Here $\GQ$ and $K$ determine the bulk and distortion energy density scales respectively. For simplicity we have adopted the one-elastic constant approximation in the distortion free energy (Eq.~\ref{eq:distortion}).

The tensor $\mathbf{M}^{(d)}(\mathbf{B},\mathbf{Q})$ is defined for an arbitrary tensor $\mathbf{B}$ as
\begin{align}
  \mathbf{M}^{(d)}(\mathbf{B},\mathbf{Q}) &= \frac{2}{d}\xi \mathbf{B}+\xi\lbrace \mathbf{B}\cdot\mathbf{Q}+\mathbf{Q}\cdot\mathbf{B}{-\frac{2}{d}\mathbf{I}\Tr{\tsr{Q\cdot B}}} \rbrace\notag\\
  & \quad - 2\xi \mathbf{Q}\Tr{\tsr{Q\cdot B}}\;,
  \label{eq:new tensor}
\end{align}
where
\begin{equation}
  \xi = \frac{Sd}{(d-2)S+2}\lambda\;.
  \label{eq:LE_compare}
\end{equation}
Here $\lambda$ is the Leslie-Ericksen flow aligning parameter, which specifies how the nematic director responds to a shear flow: $|\lambda| > 1$ corresponds to flow-aligning nematics and $|\lambda| < 1$ corresponds to the flow-tumbling regime. (See Appendix A.)

The stress tensor $\bm\Sigma$ in \eqref{eq:navier_stokes} is the sum of passive liquid-crystal and active contributions, $\bm\Sigma=\bm\Sigma^{Q} +\bm\Sigma^a$. The passive part of the stress tensor is given by
\begin{equation}
  \bm\Sigma^Q =  2[\mathbf{Q\cdot H}]^{A} -\mathbf{M}^{(d)}(\mathbf{H},\mathbf{Q})-\bm\nabla \mathbf{Q}:\frac{\delta F}{\delta \bm\nabla \mathbf{Q}}\;.
  \label{eq:Q_stress}
\end{equation}
In an active nematic there is an additional active stress contribution that arises from the dipolar forces exerted by active particles on their environment. This active stress is $\bm\Sigma^{a} = \alpha \mathbf{Q} $, where $\alpha > 0 $ describes contractile stresses and $\alpha < 0 $ extensile stresses.  In the passive limit $\alpha\to 0$, the equations just described reduce to those of a passive liquid-crystal.

So far, the model that we have presented encompasses both of the numerical studies performed.  We now outline the specific choices for parameter values and dimensionality made in each of the two numerical studies separately, and discuss how these choices affects the form of the equations.  The main difference between the two variations will be the presence of higher order coupling terms in Model II, both in the liquid-crystal stress and in the coupling between orientational order and velocity gradients.

\textbf{Model I (MI):} This model describes a $D=2$ dimensional nematic sheet with a $d=2$ dimensional nematic order parameter.  In this case the symmetric second rank tensor $\mathbf{Q}$ has only two independent components and ${\rm Tr}[\mathbf{Q}^3]=0$ identically.  In Model I the mean-field free energy (\eqref{eq:landau}) has coefficients $A=\frac{1-\Gamma}{2}$ and $C = \Gamma$, where $\Gamma$ is a dimensionless parameter that controls the continuous transition from an isotropic to a nematic state, with the transition occurring at $\Gamma=1$.  The second term in \eqref{eq:new tensor} is also identically zero; the third term is of a higher order in $\mathbf{Q}$ and can safely be neglected~\cite{Olmsted:2008}, so that $\mathbf{M}^{(2)}(\mathbf{B},\mathbf{Q})= \xi \mathbf{B}$.  We also exclude the last term in \eqref{eq:Q_stress} in Model I. We assume a constant density $\rho = 1$ for which the Reynolds number $Re = \rho V \lQ / \eta = 1$, where the velocity scale is $V = \lQ G_Q / \eta$\footnote{Note that the typical activity-induced velocity scale is in the range $V = 1 \to 10$ (see \figref{fig:spatial_avg}b or \eqref{eq:scaling_estimate}), meaning that the effective Reynolds number is in the range ${\rm Re} = 1\to10$. These values are still small enough to ensure that any turbulence is activity driven (rather than inertial) in nature.}.  We choose $\Gamma=2$ such that the system is deep in the nematic state, with $S_0=0.78$. According to \eqref{eq:LE_compare}, the system will be in the flow-aligning regime if $|\xi|>0.78$ and in the flow-tumbling regime for $|\xi|<0.78$. All results shown below for Model I correspond to $\xi = \pm 0.1$ (flow-tumbling regime).  We also choose $\xi>0$ for extensile systems ($\alpha<0$) and $\xi<0$ for contractile systems ($\alpha>0$) to guarantee $\alpha \xi < 0$, a condition that is required in to observe the initial flow instability in the ordered state~\cite{luca-RS}.

\textbf{Model II (MII):} This model considers a $D=2$ dimensional layer of nematic liquid crystal described by the full $d=3$ Landau free energy given in \eqref{eq:landau}, thereby in principle allowing the director to explore all $d=3$ dimensions. However it still neglects all spatial variations in the direction of the layer thickness, so taking $D=2$ as noted above.  In this case the free energy in \eqref{eq:landau} sets $A=1-\frac{\Gamma}{3}$, $B = -\Gamma$ and $C = \Gamma$, yielding a first order isotropic-nematic transition at $\Gamma=2.7$.  In the following we choose $\Gamma=3$ which places us at the spinodal stability limit of the isotropic phase and well within the nematic state, with $S_0 = 0.6$.  According to \eqref{eq:LE_compare}, the system will be in the flow-aligning regime for $|\xi| > 0.6$ and in the flow-tumbling regime for $|\xi| < 0.6$. In all simulations using Model II we have fixed $\xi =0.7$, corresponding to a flow-aligning system. We consider only extensile systems with this model, \ie values of $\alpha<0$.  Finally, in MII we take the limit of zero Reynolds number by setting $\rho = 0$.

\begin{table}
  \def\arraystretch{1.5}
  \resizebox{\linewidth}{!}{%
    \begin{tabular}{ l  l  l  l }
      \hline
      \textbf{parameter} & \textbf{description} &  \textbf{dimensions} \\
      \hline
      $\alpha$ & activity & $[\sigma] $ \\
      $K$ & Frank constant & $[\sigma][L]^2$ ($=1$ in MI) \\
      $\GQ$ & energy density scale  & $[\sigma]$ ($=1$ in MI, MII) \\
      $\gamma$ & rotational viscosity  & $[\sigma][T]$ ($=1$ in MII) \\
      $\xi$ & alignment param. & $[1]$ \\
      $\Gamma$ & IN control param. & $[1]$ \\
      $\eta$ & solvent viscosity  & $[\sigma][T]$ ($=1$ in MI)\\
      $\rho$ & solvent density  & $[M][L]^{-d}$\\
      $L_x=L_y=L$ & box size & $[L]$ ($=1$ in MII) \\
      \hline
    \end{tabular}
  }
  \caption{Summary of the various model parameters and their dimensions.  The choices for mass $[M]$ (or equivalently stress $[\sigma]= [M][L]^{d-2}[T]^{-2}$), length $[L]$ and time $[T]$ in each model are also indicated.}
  \label{tbl:params}
\end{table}

The full list of nine parameters (for both models) is given in Table~\ref{tbl:params}.  We are free to choose units of mass $[M]$, length $[L]$ and time $[T]$, or equivalently of stress $[\sigma]= [M][L]^{d-2}[T]^{-2}$, length $[L]$ and time $[T]$, and we have noted in Table~\ref{tbl:params} which quantities we chose to set equal to unity in each of the two studies.  This leaves six dimensionless groupings that we summarize in Table~\ref{tbl:params_dimensionless}, three of which are fixed throughout.  Therefore even though we choose our units differently in the two different simulation studies, all results are presented and compared in a consistent adimensional way between the two models. We choose parameters that produce flow-tumbling behaviour in Model I and flow-aligning in Model II.  Note that due to differences in parameter selections, the linear instability thresholds in the two models differ by a factor $O(10^3)$ ($\alpha_c/G_Q = 0.3 \to 0.4$ in Model I \cite{luca-RS} and $\alpha_c/G_Q = 4 \times 10^{-5} \to 4 \times 10^{-4}$ in Model II \cite{Hemingway:2015}). Accordingly, the onset of the turbulent regime in each model is separated by a similar factor, requiring us to explore different ranges of dimensionless activity, as noted in Table~\ref{tbl:params_dimensionless}. In particular, in Model II we explore the transition from small to large activities, whereas Model I focuses on larger activities still (\ie deeper into the regime of fully developed active turbulence).

\begin{table}
  \def\arraystretch{1.5}
  \resizebox{\linewidth}{!}{%
    \begin{tabular}{ l  p{3.1cm}  l  l  l }
      \hline
      \textbf{parameter} & \textbf{description} &  \textbf{MI value} & \textbf{MII value} \\ \hline
      \multicolumn{4}{c}{\it varied parameters}\\ \hline
      $\alpha / \GQ$ & dimensionless activity & $20 \to 10^3$ & $0.05 \to 12.8$\\
      $\frac{K}{L^2 G_Q} = \left(\frac{\ell_Q}{L}\right)^2$ & ratio of micro- to macroscopic length scales & $6.1 \times 10^{-5}$ & $2 \times 10^{-6} \to 10^{-5}$ \\
      $\gamma/\eta$ & ratio of viscosities  & $10 \to 40$ & 0.567\\ \hline
      \multicolumn{4}{c}{\it fixed parameters}\\ \hline
      $\xi$ & alignment param. & $\pm0.1$ & 0.7\\
      $\Gamma$ & IN control param. & 2 & 3 \\
      ${\rm Re} = \frac{\rho \lQ V}{\eta}$ & Reynolds number & $1$ & 0\\ \hline
    \end{tabular}
  }
  \caption{Summary of the dimensionless parameters and their values in both models. The velocity scale $V = \lQ G_Q / \eta$ in our units.}
  \label{tbl:params_dimensionless}
\end{table}

\subsection{Numerical details}

In order to demonstrate the robustness of our results with respect to numerical implementation, we use two independent codes (one for each model). In each case we perform simulations in a square box of side $L$ with biperiodic boundary conditions.  The $\tsr{Q}$ dynamics in Model I is time-integrated numerically on a square grid of $128^2$ points using a fourth order Runge-Kutta method, with a timestep $\Delta t = 10^{-3}$. Gradients of $\mathbf{Q}$ are computed using a finite difference scheme.  In Model II $\tsr{Q}$ is integrated numerically using a Euler time-stepping scheme of timestep in the range $\Delta t = 10^{-4} \to 10^{-2}$ on a grid of $256^2 \to 2048^2$ points (dependent on the magnitude of activity) and gradients of $\mathbf{Q}$ are treated using a semi-implicit Fourier method. In Model II, the velocities are determined instantaneously from the force balance equation, which we solve in Fourier space using a stream function formulation. In Model I, we integrate the Navier-Stokes equation (\eqref{eq:navier_stokes}) with the same scheme used for the order parameter equation to obtain the velocity at every time step.  We have verified that our results are quantitatively unchanged upon decreasing the timestep or grid spacing.  Both simulations were initialised with a uniform director field orientated within the $x-y$ plane, subject to a small sinusoidal perturbation of magnitude $\sim10^{-5}$.

\section{Characteristic length scales}
\label{sec:char_lengths}

Irrespective of the specific details of the model used, we expect the
resulting dynamics of the active nematic to be controlled by the
interplay of key length- and time scales that govern the basic
physics.

\subsection{Model length scales}
\label{sec:model_lengths}

An inspection of the hydrodynamic equations and model geometry reveals three underlying length scales.  The first is simply the system size $L$. The second arises from balancing the bulk and elastic-distortion free energy terms in \eqsref{eq:landau} and \ref{eq:distortion}, to obtain the equilibrium nematic persistence length, which deep in the nematic state is given by \footnote{In Model I the isotropic-nematic transition is continuous and the equilibrium nematic correlation length given by $\lQ=\sqrt{K/|A|}=\sqrt{2 K/|G_Q(1-\Gamma)|}$ diverges at the transition. Deep in the nematic state where $\Gamma\gg1$ we can approximate $\lQ\sim\sqrt{K/G_Q}$.  }
\begin{equation}
  \lQ = \sqrt{\frac{K}{\GQ}}.
  \label{eq:lQ}
\end{equation}
This is the length scale over which spatial correlations in the nematic
field decay  deep in the nematic phase, where it is
proportional to the defect core radius. The third lengthscale arises
by balancing the elastic stress $\sim K/\ell^2$ associated with a
deformation over a length $\ell$ with the active stress scale
$\sim|\alpha|$, to give the active length scale
\begin{equation}
  \la = \sqrt{\frac{K}{|\alpha|}}\;.
  \label{eq:la}
\end{equation}
To guarantee that any physics on these lengthscales $\lQ,\la$ is not contaminated by finite size effects, we focus on the regime in which $\lQ \ll L$ and $\la\ll L$.

Alternatively, from a dynamical viewpoint one might consider the system to be controlled by two timescales: the passive structural relaxation time $\tau_p = \gamma \ell^2/K$, which controls the relaxation of a distortion to the nematic order on a length scale $\ell$, and the active time scale $\tau_{\alpha} = \eta/|\alpha|$, which controls the relative rates of injection of active stresses and stress decay via viscous dissipation.  The length scale that results when these timescales are equated is then
\begin{equation}
  \ell_\tau = \sqrt{\frac{K  \eta}{\alpha \gamma}} = \la \sqrt{\frac{\eta}{\gamma}}.
\label{eq:ltau}
\end{equation}

\subsection{Emergent length scales}
\label{sec:emergent_lengths}

The length scales discussed above were motivated by simple dimensional analysis of the model parameters and flow geometry.  In our numerical simulations, we find that (for a high enough level of activity) an initially homogeneous state gives way to a spatio-temporally complicated state with defects in the nematic director field, and associated local flows in the velocity field, as found earlier by several authors~\cite{Giomi:2013,Thampi:2013} and shown in the snapshots of Figs.~\ref{fig:corr_length_MII} and \ref{fig:corr_length_MI}.  An important aim of the present work is to elucidate how the length scales associated with these emergent structures  depend on the underlying model length scales just discussed.  We denote these emergent length scales by the common symbol $\ell^*$, but in fact there are multiple possible scales that we might choose to characterize the spatio-temporal dynamics, as we now describe.

\begin{figure}[t]
  \centering
  \includegraphics[width=\columnwidth]{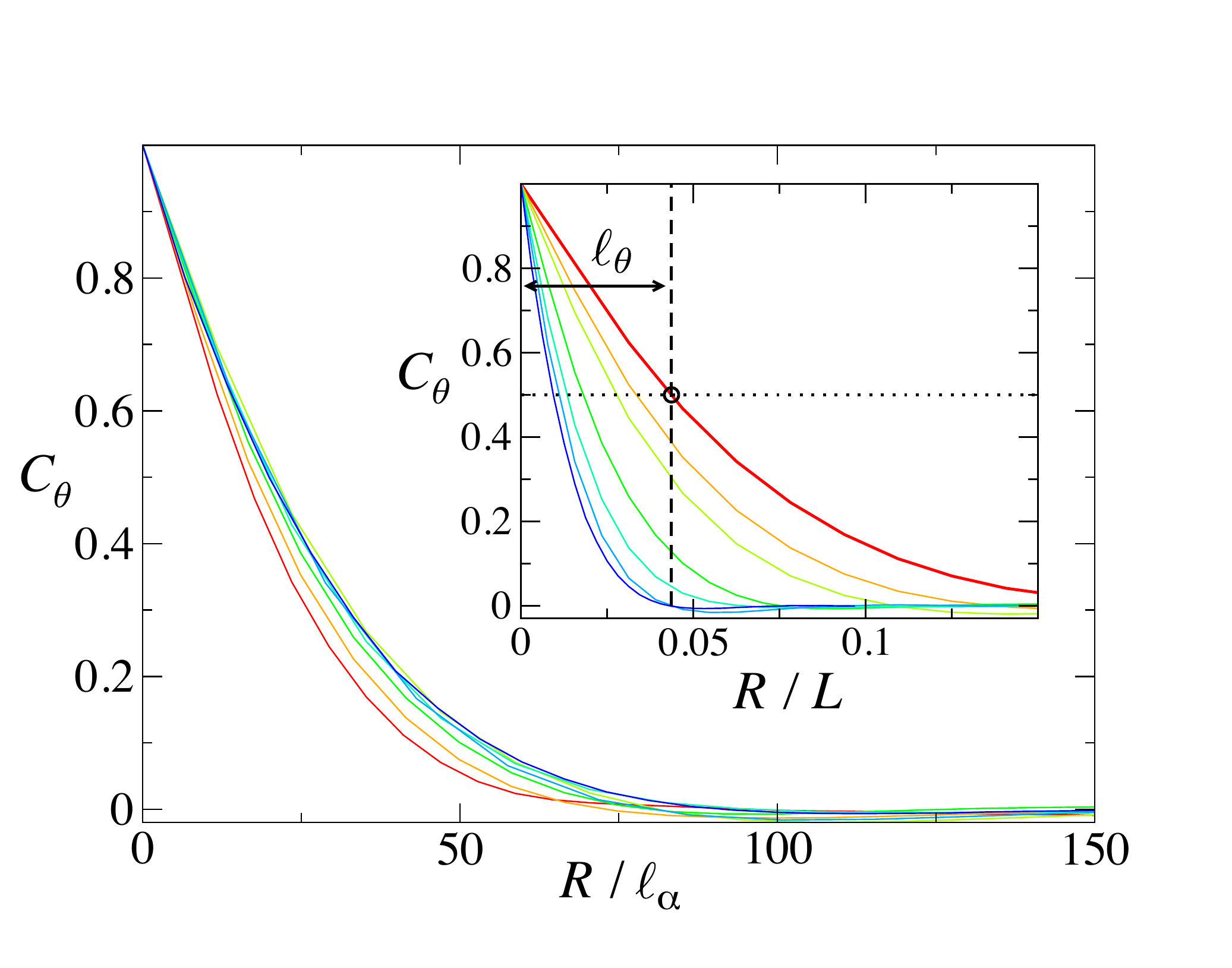}
  \vspace{-20pt}
  \caption{Nematic correlation function $C_\theta$ defined in \eqref{eq:corr_director} obtained from Model II for an extensile system in the regime of spatio-temporally chaotic behavior for $(\lQ/L)^2 = 5\times10^{-6}$ and activities in the range $\alpha/\GQ = -0.4$ (red) to $-12.8$ (blue). Inset: Unscaled data, demonstrating our definition $C_\theta(\lqq) = \sfrac{1}{2}$. Main: the same data collapse onto a single curve when rescaled by the active length $\la$.  }
  \label{fig:corr_director}
\end{figure}

\begin{itemize}
\item \textbf{Mean defect separation $\ell_d$:}
We define the mean defect separation
    \begin{equation}
      \ld = 1/\sqrt{n_d}\;,
      \label{eq:defect_spacing_length}
    \end{equation}
    where $n_d$ is the areal density of defects, calculated by
    adapting the defect tracking method of Ref.~\cite{Huterer:2005}.

  \item \textbf{Director correlation length $\lqq$:} The normalised director correlation function defined as
    \begin{equation}
      C_\theta(R) = \frac{2 \langle \vtr{n}(\mathbf{R})\cdot\vtr{n}(\mathbf{0})\rangle - 1}{2 \langle \vtr{n}(\mathbf{0})\cdot\vtr{n}(\mathbf{0})\rangle - 1}\;.
      \label{eq:corr_director}
    \end{equation}
    This characterizes the probability that two director orientations a
    distance $R$ apart are the same (respecting the fact that $\vtr{n}
    \to -\vtr{n}$ are equivalent for a nematic).  Here and throughout,
    the angular brackets $\langle \cdot \rangle$ indicate an average
    over space and time. We then choose $\lqq$ to be the length at
    which $C_\theta\left(\lqq\right) = \sfrac{1}{2}$, as in
    \figref{fig:corr_director} (inset).

  \item \textbf{Velocity correlation length $\lv$:} Analogously, the
    velocity correlation function,
    \begin{equation}
      C_v(R) = \frac{ \langle \vtr{v}(\mathbf{R})\cdot\vtr{v}(\mathbf{0})\rangle}{ \langle \vtr{v}(\mathbf{0})\cdot\vtr{v}(\mathbf{0})\rangle}\;,
      \label{eq:corr_vel}
    \end{equation}
    defines the velocity correlation length $\lv$ according to $C_v(\lv) =
    \sfrac{1}{2}$.

  \item \textbf{Vorticity correlation length $\lw$:}
    Finally, we define the correlation function for the local vorticity, $\Omega = \partial_x v_y - \partial_y v_x$, as
    \begin{equation}
          C_\Omega(R) = \frac{ \langle {\Omega}(\mathbf{R}){\Omega}(\mathbf{0})\rangle}{ \langle {\Omega}(\mathbf{0}){\Omega}(\mathbf{0})\rangle}\;,
      \label{eq:corr_vor}
    \end{equation}
 and define the vorticity correlation length $\lw$ by $C_\Omega(\lw) = \sfrac{1}{2}$.
\end{itemize}

\subsection{Scaling hypothesis}
\label{sec:scaling_general}

Simple dimensional analysis based on the model length scales discussed in Sec.~\ref{sec:model_lengths}  suggests that the length scales $\ell^*$ of Sec.~\ref{sec:emergent_lengths} characterizing the emergent structures in the fluid (whether $\ld,\lqq,\lv$ or $\lw$) should obey a simple scaling relation of the form
\begin{equation}
  \frac{ \ell^*}{\lQ} = F^*\left(\frac{\la}{\lQ},\frac{L}{\lQ}\right)\;,
  \label{eq:scaling_active_length}
\end{equation}
where $F^*$ is a general scaling function.

\begin{figure*}[t]
  \centering
  \includegraphics[width=\textwidth]{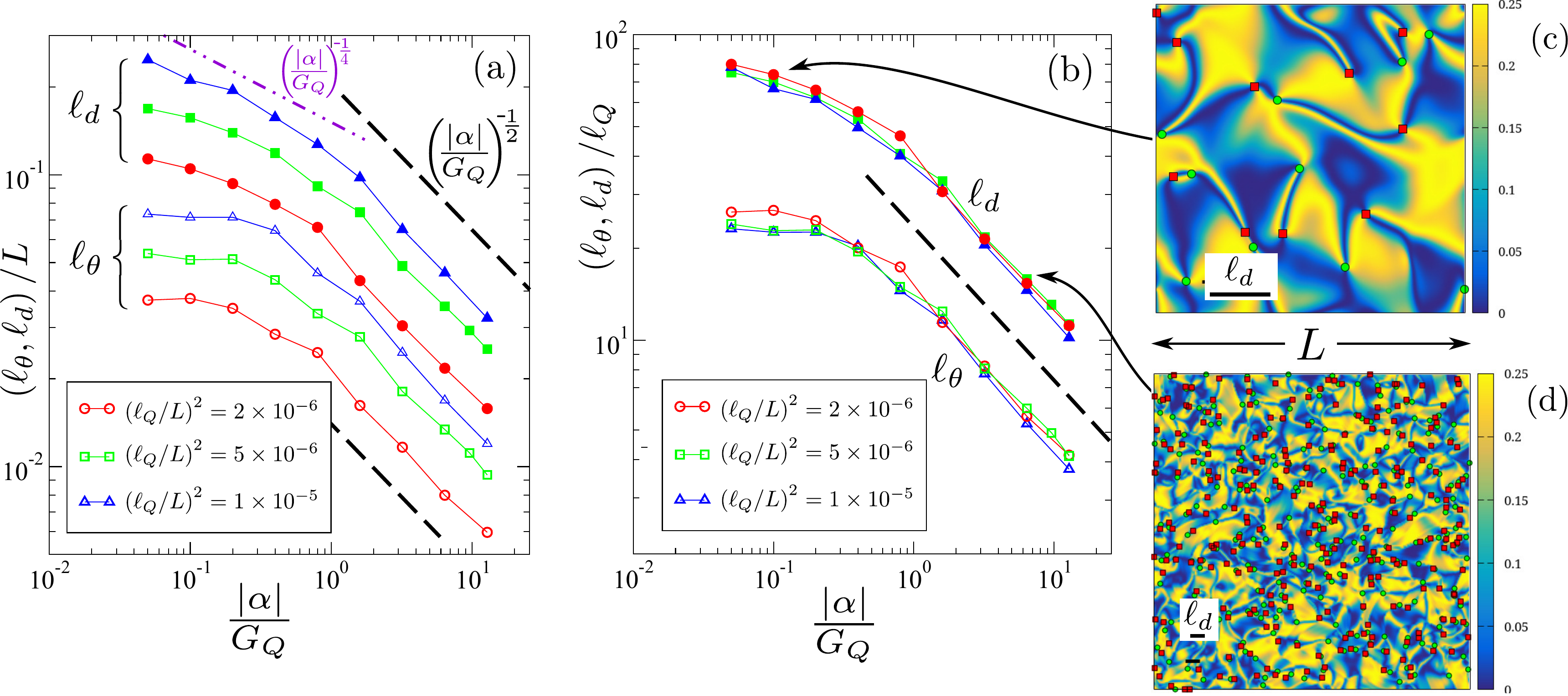}
  \vspace{-15pt}
  \caption{Results from Model II for the nematic correlation length $\lqq$ (empty symbols) and defect  spacing $\ld$ (filled symbols) as functions of the dimensionless activity $|\alpha| / \GQ$ for an extensile nematic ($\alpha < 0$).  (a) Lengthscales vs $|\alpha|$ for various values of the microscopic correlation length: $(\lQ/L)^2 = 2 \times10^{-6}$ (red circles), $5\times10^{-6}$ (green squares), and $1 \times10^{-5}$ (blue triangles). The remaining parameter values are given in Table~\ref{tbl:params_dimensionless}. At small activity we see saturation due to finite size effects. (b) The curves collapse when $\lqq$ and $\ld$ are rescaled by $\lQ$. In both frames the black dashed lines show $(|\alpha|/G_Q)^{-1/2}$. In Fig.~\ref{fig:corr_length_MII}a we also mark the power law $(|\alpha|/G_Q)^{-1/4}$ obtained by Thampi \etal as a purple dot-dashed line. (c,d) Representative snapshots of $(n_x n_y)^2$  for (c) $|\alpha|/\GQ = 0.1$ and (d) $|\alpha|/\GQ = 6.4$. We set $(\lQ/L)^2 = 1 \times 10^{-5}$ in both snapshots. Defects of topological charge $\pm1/2$ are identified by green dots (+) and red squares (-). For videos see supplementary material.
 }
  \label{fig:corr_length_MII}
\end{figure*}

Previous simulation studies\cite{Marenduzzo:2007,Fielding:2011,Hemingway:2015,Giomi:2014,Thampi:2014a} have shown that all characteristic length scales, denoted generically by $\ell^*$, decrease with increasing activity $|\alpha|$. At low activity, typically just a few defects are seen in the simulation box, as in the snapshots in Figs.~\ref{fig:corr_length_MII}c and ~\ref{fig:corr_length_MI}b.  At higher activity one obtains a state of fully developed turbulence with a much higher density of defects (Figs.~\ref{fig:corr_length_MII}d and \ref{fig:corr_length_MI}c).  In this highly turbulent regime we expect the emergent length scale $\ell^*$ to become much smaller than, and therefore independent of, the system size $L$.  The above scaling form is then accordingly expected to reduce to
\begin{equation}
 \frac{ \ell^*}{\lQ} = F^*\left(\frac{\la}{\lQ}\right)\;.
  \label{eq:scaling_active_lengthb}
\end{equation}
In our simulations all scaling law measurements are taken safely within this regime of fully developed turbulence, such that the emergent length scales are free of finite size effects. We also explicitly demonstrate that finite-size effects indeed return when $\la / L$ is no longer small, as illustrated by the snapshot of Fig.~\ref{fig:corr_length_MII}c.

It is also worth noting that at extremely large activities the defect density could in principle become so large that the defect spacing approaches the microscopic length scale $\lQ$. In this regime we would expect $\ell^*$ to be unable to decrease further upon any additional increase in activity, and so to saturate.  However our simulations do not reach this limit and the inequality $\lQ < \ell^* < L$ is always respected.

\subsection{Form of the scaling function}
\label{sec:scaling_specific}

Having proposed the existence of a scaling function in Eqn.~\ref{eq:scaling_active_lengthb}, we now consider possible specific forms for this functional dependence of $\ell^*$ on the model parameters. Conflicting scaling laws for $\ell^*$ have been proposed in the existing literature~\cite{Thampi:2014a,thampi-RS, Giomi:2014}.  While all of these studies agree that $\ell^* \propto K^{1/2}\propto \lQ$, there remains an apparent discrepancy over the scaling of $\ell^*$ with the activity.

Using Model II, Thampi \textit{et al.}~\cite{Thampi:2014a,thampi-RS} have proposed that $\ell^* \propto \alpha^{-1/4}$, which would correspond to $F^*$ in Eqn.~\ref{eq:scaling_active_lengthb} having a square root dependence on its first argument. In contrast, using Model I, Giomi~\cite{Giomi:2014} suggested the relation $\ell^* \propto \alpha^{-1/2}$, which would correspond to a linear dependence of $F^*$ on its first argument.
A possible  origin of this discrepancy is the differing dimensionality of the order parameter $\mathbf{Q}$ between the two studies: while both have $D=2$, Refs.~\cite{Thampi:2014a,thampi-RS} had $d = 3$, whereas Ref.~\cite{Giomi:2014} had $d = 2$. This motivates us to compare numerical results for both $d = 2, 3$ within a single study.  However our results below will rule out differences in $d$ as a source of discrepancy.  Another potential reason could be that the two studies in fact explored different parameter regimes given the high dimensionality of the parameter space in these models.  Therefore in order to ascertain the generality of these scaling laws, we systematically explore wide ranges for the three relevant adimensional  parameters $\left(\alpha/\GQ, \lQ/L, \gamma/\eta \right)$ for both models.  Our results will show that both forms suggested by the earlier studies can indeed apply, each in a different regime: one in the regime of fully developed active turbulence, the other when the system size plays a non-trivial role.

\section{Results}
\label{sec:results}

We now present the results of our simulations. We focus on the regime of fully developed turbulence, corresponding to activity large enough to avoid finite system-size effects ($\ell_\alpha<L$) and yet small enough to avoid saturation of the defect spacing at the microscopic length ($\ell_d>\ell_Q$). We systematically explore the functional dependence of the emergent correlation lengths defined in \secref{sec:emergent_lengths} on the model parameters.  Specifically in Model I we vary the activity, $\alpha/\GQ$, and viscosity ratio, $\gamma/\eta$, keeping all other parameters fixed to the values in Table \ref{tbl:params_dimensionless}. In Model II we vary the activity $\alpha/\GQ$ and the nematic persistence length $\lQ/L$, with all other parameters fixed to the values in Table \ref{tbl:params_dimensionless}.  We will show that in the region of fully developed active turbulence all of the emergent length scales defined above scale with the active length $\la\sim|\alpha|^{-1/2}$, in both models.  We will additionally demonstrate that a weaker exponent might be obtained in the regime of less well developed turbulence, where the typical size of the emergent structures is an appreciable fraction of the box size.

\subsection{Correlation lengths}

\begin{figure}[t]
  \centering
  \includegraphics[width=\columnwidth]{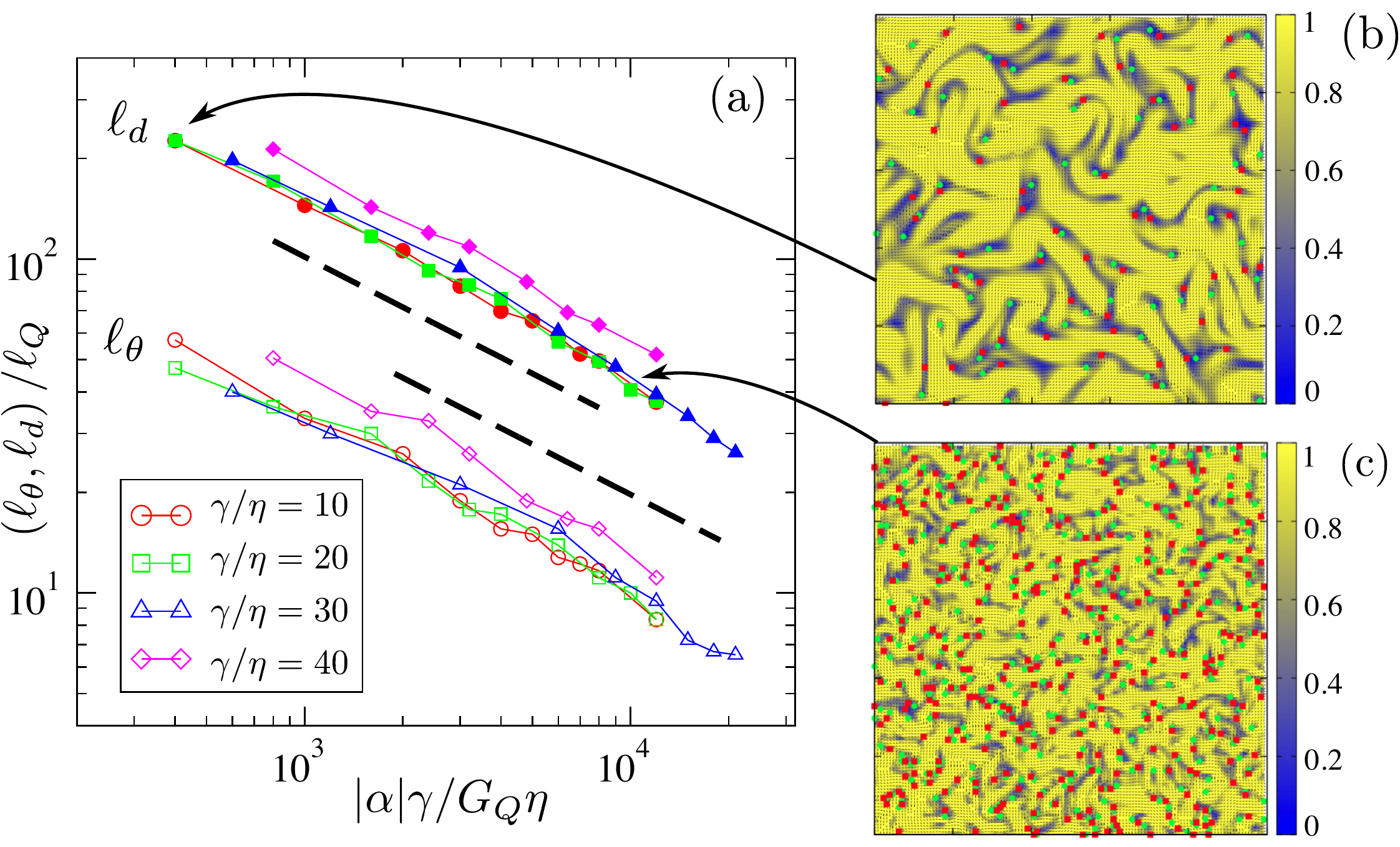}
  \vspace{-10pt}
  \caption{Results from Model I for the nematic correlation length $\lqq$ (empty symbols) and defect spacing $\ld$ (filled symbols) as functions of the dimensionless activity $|\alpha| \gamma/ \GQ\eta$ for an extensile nematic ($\alpha < 0$).  (a) Length scales vs $|\alpha|$ for  various values of the viscosity ratio: $\gamma/\eta = 10$ (red circles), $20$ (green squares), $30$ (blue triangles), and $40$ (magenta diamonds). The values of the other parameters are given in Table~\ref{tbl:params_dimensionless}.
  The black dashed lines denote a slope of $-1/2$. (b,c) Representative snapshots of the alignment tensor for $\eta/\gamma = 20$ in (b) the low activity regime ($|\alpha|/G_Q=20 $) with low defect density and (c) the high activity regime ($|\alpha|/G_Q=100 $) with high defect density. The color scale represents the magnitude $S$ of the order parameter and the black lines denote the local orientation of the director field. Topological defects with charge $\pm1/2$ are shown as green dots (+)  and  red squares (-).  }
  \label{fig:corr_length_MI}
\end{figure}

In this section, we present our results for the correlation lengths
defined in Sec.~\ref{sec:emergent_lengths}. Our main focus will be on
an extensile nematic, corresponding to $\alpha
< 0$. We shall briefly discuss the contractile case at the end of this section.

\subsubsection{Extensile active matter}

\textit{Orientational correlations.} We begin by considering correlations in the nematic order parameter $\mathbf{Q}$.  Figs.~\ref{fig:corr_length_MII}a and \ref{fig:corr_length_MI}a shows  the director correlation length $\lqq$ and the defect spacing $\ld$ as obtained from Model II and Model I, respectively.  For sufficiently large activity $\alpha$, we find that in both models both lengths obey a clear scaling law $\ld, \lqq \sim \left(\alpha/\GQ\right)^{-1/2}$ (black dashed lines).  Note that the defect spacing correlation length $\ld$ is consistently larger than $\lqq$ by a factor $\sim 2-3$. This is to be expected as correlations at the halfway point between two defects ($\ld/2$) should be similar to those at $\lqq$.

At smaller activities (\ie for $|\alpha|/\GQ \lesssim 1$) the data obtained with Model II show a saturation in the power law (leftmost data points in Fig.~\ref{fig:corr_length_MII}a). This can be attributed to that fact that the length scale of nematic structure now spans an appreciable fraction of the system size, as seen in the snapshots of Fig.~\ref{fig:corr_length_MII}c. It is possible that fitting a power law in this saturation regime could result in a less negative exponent than the $-1/2$ found in the regime of fully developed turbulence.  Indeed we find that the scaling $|\alpha|^{-1/4}$ suggested by Thampi \textit{et al.}  (purple dashed dotted line in Fig.~\ref{fig:corr_length_MII}a) matches our data reasonably well in this regime.

The data in Fig.~\ref{fig:corr_length_MII}a also suggests that both $\ld$ and $\lqq$ scale linearly with $\lQ$. We verify this scaling explicitly in Fig.~\ref{fig:corr_length_MII}b by plotting $\ld / \lQ$ and $\lqq/\lQ$ against activity. The data for various values of $\lQ$ collapse neatly onto a single curve, demonstrating a clear linear relation between both correlation lengths and $\lQ$.

The data obtained with Model I shown in Fig.~\ref{fig:corr_length_MI}a focus on large activities and verify that in this regime the scaling of both $\ld$ and $\lqq$ with $ \left(|\alpha| / \GQ\right)^{-1/2}$ holds regardless of the model used. (They do not probe the saturation with system size seen at lower activities in Model II.) Data obtained for different values of the viscosity ratio $\gamma / \eta$ can be collapsed when plotted  as shown in Fig.~\ref{fig:corr_length_MI}b, suggesting $\ell^*/\lQ\sim\ell_\tau/\lQ= \left[|\alpha|\gamma/(\GQ\eta)\right]^{-1/2} $, although the range of variation of the viscosity ratio is not sufficient to provide convincing evidence of scaling.

\begin{figure}[t]
  \centering
  \includegraphics[width=\columnwidth]{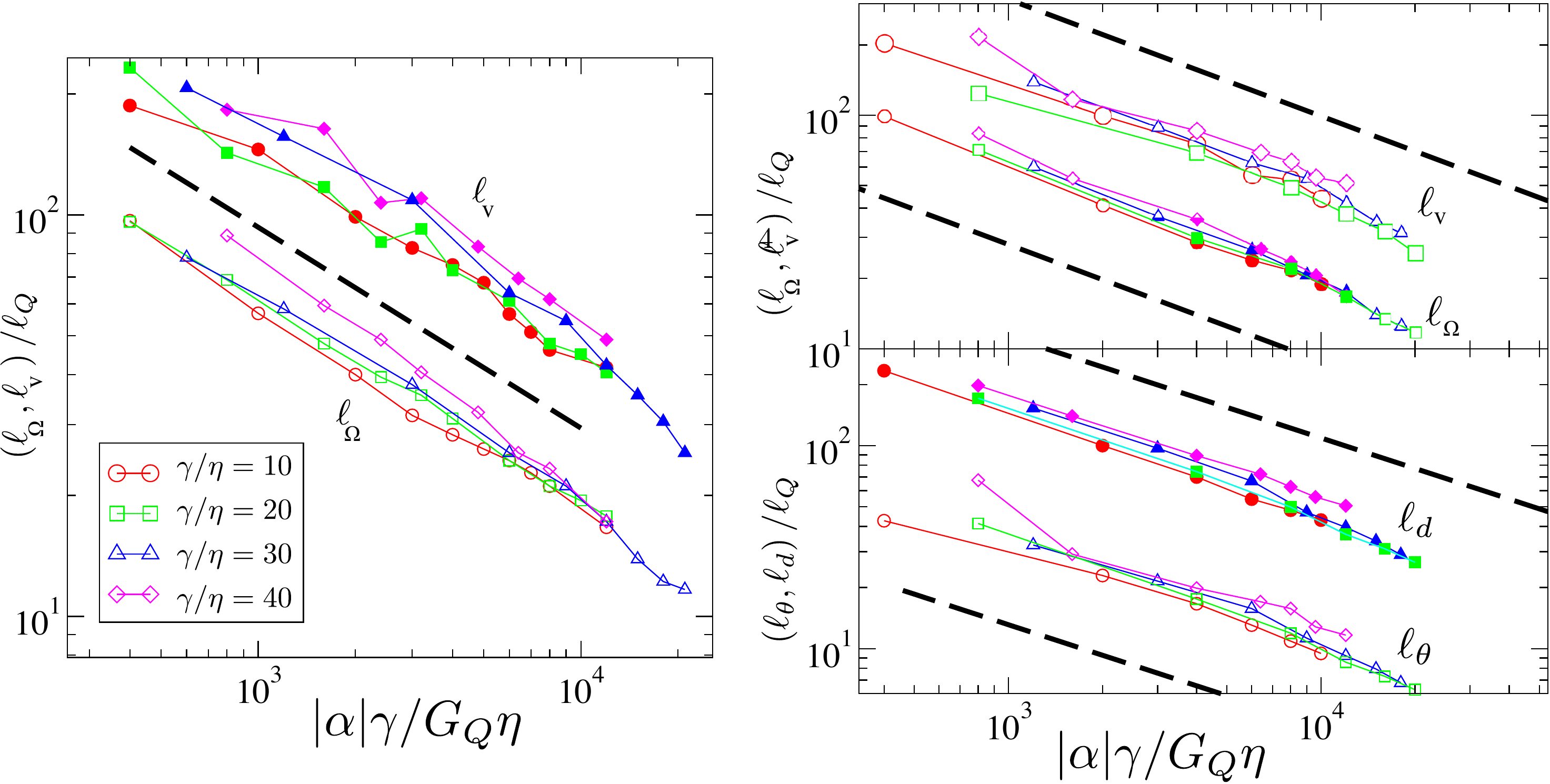}
  \caption{Velocity ($\ell_v$, filled symbols) and vorticity ($\ell_\Omega$, empty symbols) correlation lengths, normalized by $\lQ$ for (a) an extensile  ($\alpha < 0$) and (b) a contractile ($\alpha > 0$) system. We explore several  values of the viscosity ratio: $\gamma/\eta = 10$ (red circles), $20$ (green squares), $30$ (blue triangles), and $40$ (magenta diamonds). Frame (c) shows the defect spacing $\ell_d$ (filled symbols) and the director correlation length $\ell_\theta$ (empty symbols) for a contractile ($\alpha > 0$) active nematic as a function of $\alpha \gamma / \GQ \eta$ for the same set of values of $\gamma/\eta$. All lengths scale as $\left(\alpha \gamma / \GQ \eta\right)^{-1/2}$.  The black dashed lines represent a slope of $-1/2$.  }
  \label{fig:other_corr_lengths}
\end{figure}

Taken together, the data obtained with the two models tests the functional dependence of the two nematic correlation lengths with respect to activity and the nematic persistence length $\lQ$. Once free of the system size, we find that both obey $\ell^*/\lQ \sim \la/\lQ$.  Consistent with this scaling, replotting in Fig.~\ref{fig:corr_director} (main) the full director correlation function as a function of the rescaled coordinate $R/\la$ gives good data collapse. Additionally, the data obtained with Model I suggest a scaling $\ell^*/\lQ \sim (\la/\lQ)\sqrt{\gamma/\eta}$, but a larger range of $\gamma$ values would be needed to verify this.  Next we demonstrate that the same scaling form is observed for correlations lengths associated with the velocity field $\vtr{v}$.

{\it Velocity and vorticity correlation lengths.} Using data obtained with Model I, we explore the dependence of the velocity correlation length $\lv$ and the vorticity correlation length $\lw$ (as defined in \secref{sec:emergent_lengths}) on activity. In light of the results of the previous section, we directly plot both these lengths against the rescaled activity $|\alpha| \gamma / \GQ \eta$ (see \figref{fig:other_corr_lengths}a).  As shown previously for the orientation correlation lengths, we again observe that both $\lv$ and $\lw$  scale as $\sim \left(|\alpha| \gamma / \GQ \eta\right)^{-1/2}$, with all data sets falling approximately on a single curve.  We stress that this behavior is different from that reported in Ref.~\cite{Thampi:2014a}, where it was argued that  $\lv$ does not depend on activity, while $\lw$ scale as $\alpha^{-1/4}$.

\subsubsection{Contractile active matter. }

So far we have presented data for extensile systems, corresponding to $\alpha < 0$. However many examples of contractile active matter are found in nature, \eg suspensions of \textit{Chlamydomonas} algae\cite{Rafai2010}, or cytoskeletal actomyosin networks~\cite{Bendix2008}.  Therefore in order to further demonstrate the generality of our results, we now briefly consider the contractile case ($\alpha > 0$). Since the linear instability of the homogeneous state requires  requires  $\alpha \xi < 0$\cite{luca-RS}, for contractile systems we use $\xi \to -\xi = -0.1$. Our data, shown in \figsref{fig:other_corr_lengths}(b,c), support the idea that the defect spacing ($\ld$), director correlation length ($\lqq$), velocity ($\lv$) and vorticity ($\lw$) correlation lengths are all controlled by a single active length scale $\la \sim |\alpha|^{-1/2}$.  We caution, however, that the mapping of rod-like extensile ($\xi > 0, \alpha < 0$) onto disc-like contractile ($\xi < 0, \alpha > 0$) only holds at the linear instability level; the full non-linear dynamics may be subject to additional instabilities depending on the specific parameter values. While we do not expect that this would significantly change the scaling behaviour, we defer a full study of these effects to future work.

\subsection{Kinetic energy and enstrophy}

\begin{figure*}
  \centering
  \includegraphics[width=0.85\textwidth]{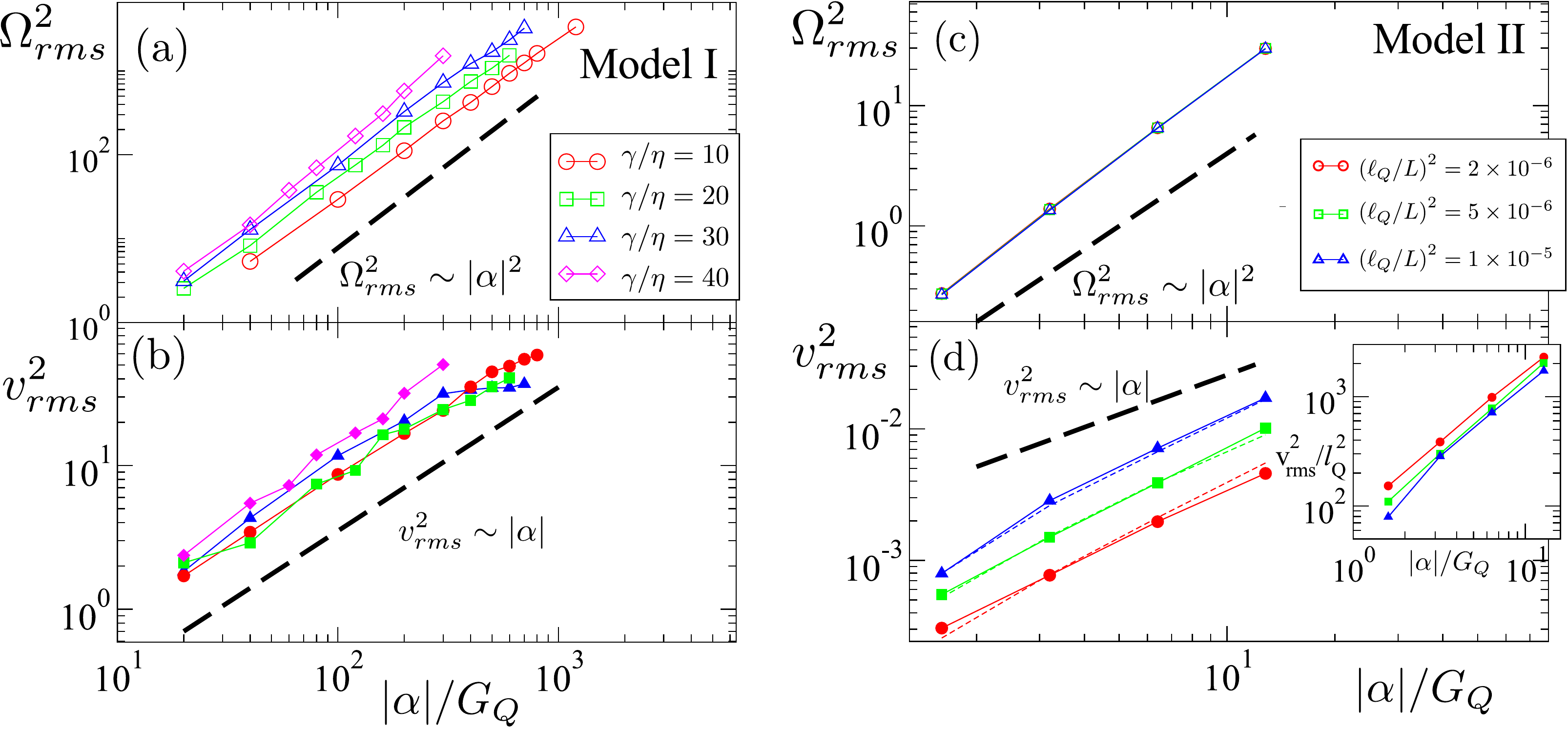}
  \caption{Scaling of kinetic energy (frames (a) and (c)) and enstrophy (frames (b) and (d)) with activity for both Models I and II. The left figure displays the results obtained from Model 1 by varying the viscosity ratio $\gamma/\eta$ as shown. The right figure displays the results obtained from Model II by varying the nematic correlation length $\lQ/L$.  The inset of frame (d) shows the scaling collapse of the kinetic energy when plotting $v_{rms}^2 / \lQ^2$ against activity. In frame (d), data is shown for two numerical resolutions: dashed lines for $N = 1024$, and solid lines for $N = 2048$.}
\label{fig:spatial_avg}
\end{figure*}

The above scaling relations were obtained using the correlation functions defined in \secref{sec:char_lengths}, which are normalised so that each function, \eg $C_v(R)$, approaches unity as the separation distance $R \to 0$. (See \figref{fig:corr_director}.) The normalization constants themselves, however, (\ie the denominators in \eqsref{eq:corr_vel} and \ref{eq:corr_vor}) also provide useful information as they are directly proportional to the mean kinetic energy and enstrophy of the system, given by
\begin{align}
  E_k=\frac{1}{2}v^2_{rms} &= \frac{1}{2}\langle \mathbf{v}(\mathbf{r})\cdot\mathbf{v}(\mathbf{r})\rangle\;,\\
  E_s=  \frac{1}{2}\Omega^2_{rms} &= \frac{1}{2}\langle \Omega(\mathbf{r})\Omega(\mathbf{r})\rangle\;,
  \label{eq:enst_kntc}
\end{align}
where the angular brackets $\langle \cdot \rangle$ again denote an average over space and time. These quantities can be obtained experimentally, for instance by using particle image velocity (PIV) to quantify the flow fields of active liquids, as done by Dunkel \etal\cite{Dunkel2013a} in suspensions of extensile \textit{B. subtilis} bacteria.  We now use our earlier findings to motivate the expected scaling relation of these flow properties with activity, and then verify our predictions with further numerical data from both models.

Using simple dimensional analysis, the characteristic velocity of  activity-induced shear flows associated with distortion of the local nematic order over a length scale $\ell^*$ can be obtained from the force balance condition (\eqref{eq:navier_stokes}) as
\begin{equation}
  v \sim |\alpha|\ell^*/\eta.
  \label{eq:scaling_estimate}
\end{equation}
Our results indicate that for sufficiently large values of activity the  physics is controlled by a single active length scale $\ell^*$, with $\ell^*\sim|\alpha|^{-1/2}$. Using this in~\eqref{eq:scaling_estimate}, we find  $v\sim|\alpha|^{1/2}$ and $<v_{rms}^2>\sim|\alpha|$.  The scaling of the vorticity can be estimated as $\omega\sim v/\ell^*$, which gives an enstrophy $\Omega_{rms}^2\sim|\alpha|^2$.

This scaling is consistent with the findings of Ref.~\cite{Giomi:2014}
in which the author examined the typical size of vortex structures in
the regime of spatio-temporal chaotic dynamics using what we refer to here as Model I and found that both the vortex size and
the defect spacing appear to scale with the active lengthscale $\la$.
Further evidence for this scaling can be found in the experiments of
Ref.~\cite{Dunkel2013a}, which found that $\Omega_{rms}^2 = v_{rms}^2
/ \left(\ell^*\right)^2$ where $\ell^*$ is the characteristic vortex
size: assuming that $\ell^* \sim \la$, this implies that
$\Omega_{rms}^2 \sim |\alpha| v_{rms}^2$ as we have argued above.  Our
proposed scaling is not, however, in agreement with the findings of
Ref.~\cite{Thampi:2013,thampi-RS}. In those studies it was found that
$v_{rms}^2 \sim |\alpha|^2$ and $\Omega_{rms}^2 \sim |\alpha|^2$, a
result that cannot seemingly be reconciled with the simple assumption
that $\Omega\sim v / \la$.

In order to appraise these conflicting scaling laws, we perform simulations with both Model I and Model II and measure the kinetic energy ($v_{rms}^2$, see \figsref{fig:spatial_avg}a, b) and enstrophy ($\Omega_{rms}^2$, \figsref{fig:spatial_avg}c, d). The data from both models clearly obey our expected scaling laws $v_{rms}^2\sim|\alpha|$ and $\Omega_{rms}^2\sim|\alpha|^2$ (black dashed lines). With Model I our choice of units means that increasing the rotational viscosity $\gamma$ is equivalent to reducing the solvent viscosity $\eta$. Our dimensional analysis in \eqref{eq:scaling_estimate} suggests that increasing $\gamma/\eta$ should increase the characteristic velocity. We indeed observe this trend in our data in \figsref{fig:other_corr_lengths}b, although simulations over a larger range of $\gamma/\eta$ would be required to determine the exact scaling. By the same analysis, we also expect that $v_{rms}^2$ should be proportional to $\ell_Q^2$ for fixed $G_Q$, since $v_{rms}^2\sim (\alpha\ell^*/\eta)^2\sim (|\alpha|G_Q/\eta^2)\ell_Q^2$.  Our data from Model II explores several values of $\lQ$, and plotting $v_{rms}^2 / \lQ^2$ against activity indeed leads to a reasonable curve collapse (\figref{fig:spatial_avg}d inset). Consistent with the findings of Giomi\cite{Giomi:2014}, we observe no appreciable dependence of $\Omega^2_{rms}$ on $\lQ$. This follows again from the scaling, $\Omega^2_{rms}\sim v_{rms}^2/\ell^{*2}\sim(\alpha/\eta)^2$.

\section{Discussion}
\label{sec:discussion}

Using two distinct continuum models that have been studied extensively within the literature, we have performed a detailed numerical study of an active nematic to examine the scaling with activity of a number of structural and hydrodynamic correlation lengths, including the mean defect spacing.  Our findings are consistent with the suggestion first put forward in Ref.~\cite{Giomi:2014} that in the regime of fully developed active turbulence defect proliferation, and the associated turbulent-like dynamics of the active nematic, are controlled by a single length scale $\la\sim|\alpha|^{-1/2}$. This is also the length scale that controls the onset of spontaneous flow instability of active films~\cite{Voituriez:2005,Marenduzzo:2007,Edwards:2009}. Our numerical data from both models show that all measures of correlation length considered scale with this length scale, for both extensile and contractile systems.

Two caveats must, however, be applied. First, for extremely large activities (\ie $\la \sim \lQ$) activity-induced deformations below the nematic persistence length $\lQ$ are expected to be suppressed.  Secondly, at  low activities, structures can form that span the system size, and correlation lengths will correspondingly saturate, (\ie $\la \sim O(L)$). We have explicitly demonstrated this system-size saturation in our simulations, a result that reconciles the apparently conflicting power law exponents previously reported in the literature.

Finally, to further support our findings, we have calculated the average kinetic energy and average enstrophy of the system, quantities that are readily obtainable from experiment. Our numerical results show that the scaling of these quantities with activity is consistent with a simple dimensional analysis based on the assumption that the physics is controlled by the single length scale $\la$. %

Our results show that the key scaling relations hold for both strictly 2D and quasi-2D models. Encouragingly, this implies that such models capture the dynamics of active nematics in a generic way, \ie independent of the specifics of the model.  How our results would compare with the equivalent fully-$3D$ simulation of an active nematic remains an interesting open question.

\section{Acknowledgment}
We thank Mark Bowick for introducing us to Ref.~\cite{Huterer:2005} (that describes a method for defect tracking) and Luca Giomi for writing the code used to study Model I. We thank them both and Mike Cates for invaluable discussions. MCM and PM were supported by the National Science Foundation through award DMR-1305184 and by the Syracuse Soft Matter Program.  EJH thanks EPSRC for a Studentship.  SMF's and EJH's research leading to these results has received funding from the European Research Council under the European Union's Seventh Framework Programme (FP7/2007-2013) / ERC grant agreement number 279365.  The authors thank the KITP at the University of California, Santa Barbara, where they were supported through National Science Foundation Grant No. NSF PHY11-25925.

\appendix

\section{Flow-aligning Parameter}

We include here a comparison between the tumbling parameter $\xi$ used here and the Leslie-Erickson (LE) tumbling parameter $\lambda$ (where $|\lambda| > 1$ corresponds to flow-aligning regime and $|\lambda| < 1$ corresponds to flow-tumbling regime). This comparison is presented in Appendix B of Ref.~\cite{Marenduzzo:2007} for the case $d=3$, but to our knowledge has not been displayed before for the case $d=2$.

In $d$ dimensions the nematic tensor $\mathbf{Q}_{ij}$ of a uniaxial nematic can be written as
\begin{equation}
  Q_{ij}=\frac{d}{(I/\Delta I)}S (n_i n_j - \frac{1}{d}\delta_{ij})\;,
  \label{eq:tensor}
\end{equation}
where $I$ and $\Delta I$ are the sum and difference, respectively of the two principal values of the moment of inertia tensor of uniaxial nematogens.  In our case we use  $I/\Delta I = 2$ that corresponds to needle-like molecules\cite{Stark:2003}.

We write the dynamical equation for the alignment tensor in $d$ dimension using the notation of Olmsted\cite{Olmsted:2008},
\begin{equation}
  \begin{split}
    \frac{D Q_{ij}}{Dt} = \Omega_{ik}Q_{kj} - Q_{ik}\Omega_{kj} + \beta_1 D_{ij} + \frac{1}{\beta_2} H_{ij} \\+ \beta_5 \lbrace Q_{ik} D_{kj} + D_{ik} Q_{kj} - \frac{2}{d} \delta_{ij} \mathbf{D}: \mathbf{Q}\rbrace \\ + \beta_6 \lbrace Q_{ik} H_{kj} + H_{ik}Q_{kj} - \frac{2}{d} \delta_{ij} \mathbf{H} : \mathbf{Q}\rbrace,
  \end{split}
  \label{eq:dynamics}
\end{equation}
where $\beta_1$, $\beta_2$, $\beta_5$ and $\beta_6$ are parameters that couple order and flow.

Substituting the expression given in \eqref{eq:tensor} for the alignment tensor into the dynamical equation \eqref{eq:dynamics}, and assuming $S$ to be constant, we  obtain an equation for the director,
\begin{equation}
  \begin{split}
    \dot{n_i} = (\bm\Omega\times \hat{\mathbf{n}})_i + \left[\frac{(I/\Delta I)^2}{2 \beta_2 (d S)^2 } + \frac{(1-\frac{2}{d})(I/\Delta I)\beta_6}{2 d S}\right] h_i \\+ \left[\frac{(I/\Delta I)\beta_1}{d S}+\left(1-\frac{2}{d}\right)\beta_5 \right] n_j D_{ij}.
  \end{split}
  \label{eq:director dynamics}
\end{equation}
Comparing \eqref{eq:director dynamics} to the Leslie-Erickson equation~\cite{Olmsted:2008},
\begin{equation}
  \dot{n_i} = (\bm\Omega \times \hat{\mathbf{n}})_i + \frac{1}{\gamma_1}h_i + \lambda n_j D_{ij},
\end{equation}
we identify the correspondence between the Olmsted coefficients $\beta_i$ and the Leslie-Erickson coefficients as
\begin{equation}
  S \lambda = \frac{(I/\Delta I)}{d} \beta_1 + \frac{(d-2)}{d}  \beta_5 S + O(S^2),
  \label{eq:LE_compare_Appendix}
\end{equation}
\begin{equation}
  \frac{S^2}{\gamma_1} = \frac{(I/\Delta I)^2}{2 \beta_2 d^2}  + \frac{(1-\frac{2}{d})(I/\Delta I)}{2d}\beta_6 S + O(S^2).
\end{equation}
Using $I/\Delta I = 2$ in \eqref{eq:LE_compare_Appendix}, we obtain
\begin{equation}
  \lambda =  \frac{2}{d} \frac{\beta_1}{S}+ \frac{(d-2)}{d} \beta_5.
\end{equation}
Finally, for the case $\beta_1=\beta_5 = \xi$, we find
\begin{align}
  \xi =
  \begin{cases}
    \lambda S & \quad \text{for } d=2\\
    \frac{3 S}{S + 2} \lambda  & \quad \text{for } d=3\\
  \end{cases}
\end{align}
The $d=3$ case was previously reported in Ref.~\cite{Marenduzzo:2007}.

\bibliography{paper_SoftMatter}

\providecommand*{\mcitethebibliography}{\thebibliography}
\csname @ifundefined\endcsname{endmcitethebibliography}
{\let\endmcitethebibliography\endthebibliography}{}
\begin{mcitethebibliography}{37}
\providecommand*{\natexlab}[1]{#1}
\providecommand*{\mciteSetBstSublistMode}[1]{}
\providecommand*{\mciteSetBstMaxWidthForm}[2]{}
\providecommand*{\mciteBstWouldAddEndPuncttrue}
  {\def\EndOfBibitem{\unskip.}}
\providecommand*{\mciteBstWouldAddEndPunctfalse}
  {\let\EndOfBibitem\relax}
\providecommand*{\mciteSetBstMidEndSepPunct}[3]{}
\providecommand*{\mciteSetBstSublistLabelBeginEnd}[3]{}
\providecommand*{\EndOfBibitem}{}
\mciteSetBstSublistMode{f}
\mciteSetBstMaxWidthForm{subitem}
{(\emph{\alph{mcitesubitemcount}})}
\mciteSetBstSublistLabelBeginEnd{\mcitemaxwidthsubitemform\space}
{\relax}{\relax}

\bibitem[Marchetti \emph{et~al.}(2013)Marchetti, Joanny, Ramaswamy, Liverpool,
  Prost, Rao, and Simha]{Marchetti:2013}
M.~C. Marchetti, J.~F. Joanny, S.~Ramaswamy, T.~B. Liverpool, J.~Prost, M.~Rao
  and R.~A. Simha, \emph{Rev. Mod. Phys.}, 2013, \textbf{85}, 1143--1189\relax
\mciteBstWouldAddEndPuncttrue
\mciteSetBstMidEndSepPunct{\mcitedefaultmidpunct}
{\mcitedefaultendpunct}{\mcitedefaultseppunct}\relax
\EndOfBibitem
\bibitem[{R. Voituriez} \emph{et~al.}(2005){R. Voituriez}, {J. F. Joanny}, and
  {J. Prost}]{Voituriez:2005}
{R. Voituriez}, {J. F. Joanny} and {J. Prost}, \emph{Europhys. Lett.}, 2005,
  \textbf{70}, 404--410\relax
\mciteBstWouldAddEndPuncttrue
\mciteSetBstMidEndSepPunct{\mcitedefaultmidpunct}
{\mcitedefaultendpunct}{\mcitedefaultseppunct}\relax
\EndOfBibitem
\bibitem[Marenduzzo \emph{et~al.}(2007)Marenduzzo, Orlandini, Cates, and
  Yeomans]{Marenduzzo:2007}
D.~Marenduzzo, E.~Orlandini, M.~E. Cates and J.~M. Yeomans, \emph{Phys. Rev.
  E}, 2007, \textbf{76}, 031921\relax
\mciteBstWouldAddEndPuncttrue
\mciteSetBstMidEndSepPunct{\mcitedefaultmidpunct}
{\mcitedefaultendpunct}{\mcitedefaultseppunct}\relax
\EndOfBibitem
\bibitem[Giomi \emph{et~al.}(2008)Giomi, Marchetti, and Liverpool]{Giomi:2008}
L.~Giomi, M.~C. Marchetti and T.~B. Liverpool, \emph{Phys. Rev. Lett.}, 2008,
  \textbf{101}, 198101\relax
\mciteBstWouldAddEndPuncttrue
\mciteSetBstMidEndSepPunct{\mcitedefaultmidpunct}
{\mcitedefaultendpunct}{\mcitedefaultseppunct}\relax
\EndOfBibitem
\bibitem[Ramaswamy \emph{et~al.}(2003)Ramaswamy, Simha, and
  Toner]{Ramaswamy:2003}
S.~Ramaswamy, R.~A. Simha and J.~Toner, \emph{EPL (Europhysics Letters)}, 2003,
  \textbf{62}, 196\relax
\mciteBstWouldAddEndPuncttrue
\mciteSetBstMidEndSepPunct{\mcitedefaultmidpunct}
{\mcitedefaultendpunct}{\mcitedefaultseppunct}\relax
\EndOfBibitem
\bibitem[Mishra and Ramaswamy(2006)]{Mishra:2006}
S.~Mishra and S.~Ramaswamy, \emph{Phys. Rev. Lett.}, 2006, \textbf{97},
  090602\relax
\mciteBstWouldAddEndPuncttrue
\mciteSetBstMidEndSepPunct{\mcitedefaultmidpunct}
{\mcitedefaultendpunct}{\mcitedefaultseppunct}\relax
\EndOfBibitem
\bibitem[Narayan \emph{et~al.}(2007)Narayan, Ramaswamy, and
  Menon]{Narayan:2007}
V.~Narayan, S.~Ramaswamy and N.~Menon, \emph{Science}, 2007, \textbf{317},
  105\relax
\mciteBstWouldAddEndPuncttrue
\mciteSetBstMidEndSepPunct{\mcitedefaultmidpunct}
{\mcitedefaultendpunct}{\mcitedefaultseppunct}\relax
\EndOfBibitem
\bibitem[Giomi \emph{et~al.}(2011)Giomi, Mahadevan, Chakraborty, and
  Hagan]{Giomi:2011}
L.~Giomi, L.~Mahadevan, B.~Chakraborty and M.~F. Hagan, \emph{Phys. Rev.
  Lett.}, 2011, \textbf{106}, 218101\relax
\mciteBstWouldAddEndPuncttrue
\mciteSetBstMidEndSepPunct{\mcitedefaultmidpunct}
{\mcitedefaultendpunct}{\mcitedefaultseppunct}\relax
\EndOfBibitem
\bibitem[Giomi \emph{et~al.}(2012)Giomi, Mahadevan, Chakraborty, and
  Hagan]{Giomi:2012}
L.~Giomi, L.~Mahadevan, B.~Chakraborty and M.~F. Hagan, \emph{Nonlinearity},
  2012, \textbf{25}, 2245\relax
\mciteBstWouldAddEndPuncttrue
\mciteSetBstMidEndSepPunct{\mcitedefaultmidpunct}
{\mcitedefaultendpunct}{\mcitedefaultseppunct}\relax
\EndOfBibitem
\bibitem[Pismen(2013)]{Pismen:2013}
L.~M. Pismen, \emph{Phys. Rev. E}, 2013, \textbf{88}, 050502\relax
\mciteBstWouldAddEndPuncttrue
\mciteSetBstMidEndSepPunct{\mcitedefaultmidpunct}
{\mcitedefaultendpunct}{\mcitedefaultseppunct}\relax
\EndOfBibitem
\bibitem[Giomi \emph{et~al.}(2013)Giomi, Bowick, Ma, and Marchetti]{Giomi:2013}
L.~Giomi, M.~J. Bowick, X.~Ma and M.~C. Marchetti, \emph{Phys. Rev. Lett.},
  2013, \textbf{110}, 228101\relax
\mciteBstWouldAddEndPuncttrue
\mciteSetBstMidEndSepPunct{\mcitedefaultmidpunct}
{\mcitedefaultendpunct}{\mcitedefaultseppunct}\relax
\EndOfBibitem
\bibitem[{Shi Xia-qing} and {Ma Yu-qiang}(2013)]{Shi:2013}
{Shi Xia-qing} and {Ma Yu-qiang}, \emph{Nat Commun}, 2013, \textbf{4},
  1218\relax
\mciteBstWouldAddEndPuncttrue
\mciteSetBstMidEndSepPunct{\mcitedefaultmidpunct}
{\mcitedefaultendpunct}{\mcitedefaultseppunct}\relax
\EndOfBibitem
\bibitem[Giomi \emph{et~al.}(2014)Giomi, Bowick, Mishra, Sknepnek, and
  {Cristina Marchetti}]{luca-RS}
L.~Giomi, M.~J. Bowick, P.~Mishra, R.~Sknepnek and M.~{Cristina Marchetti},
  \emph{Philosophical Transactions of the Royal Society of London A:
  Mathematical, Physical and Engineering Sciences}, 2014, \textbf{372},
  0365\relax
\mciteBstWouldAddEndPuncttrue
\mciteSetBstMidEndSepPunct{\mcitedefaultmidpunct}
{\mcitedefaultendpunct}{\mcitedefaultseppunct}\relax
\EndOfBibitem
\bibitem[Thampi \emph{et~al.}(2013)Thampi, Golestanian, and
  Yeomans]{Thampi:2013}
S.~P. Thampi, R.~Golestanian and J.~M. Yeomans, \emph{Phys. Rev. Lett.}, 2013,
  \textbf{111}, 118101\relax
\mciteBstWouldAddEndPuncttrue
\mciteSetBstMidEndSepPunct{\mcitedefaultmidpunct}
{\mcitedefaultendpunct}{\mcitedefaultseppunct}\relax
\EndOfBibitem
\bibitem[Thampi \emph{et~al.}(2014)Thampi, Golestanian, and
  Yeomans]{Thampi:2014a}
S.~P. Thampi, R.~Golestanian and J.~M. Yeomans, \emph{EPL (Europhysics
  Letters)}, 2014, \textbf{105}, 18001\relax
\mciteBstWouldAddEndPuncttrue
\mciteSetBstMidEndSepPunct{\mcitedefaultmidpunct}
{\mcitedefaultendpunct}{\mcitedefaultseppunct}\relax
\EndOfBibitem
\bibitem[Thampi \emph{et~al.}(2014)Thampi, Golestanian, and Yeomans]{thampi-RS}
S.~P. Thampi, R.~Golestanian and J.~M. Yeomans, \emph{Philosophical
  Transactions of the Royal Society of London A: Mathematical, Physical and
  Engineering Sciences}, 2014, \textbf{372}, 0366\relax
\mciteBstWouldAddEndPuncttrue
\mciteSetBstMidEndSepPunct{\mcitedefaultmidpunct}
{\mcitedefaultendpunct}{\mcitedefaultseppunct}\relax
\EndOfBibitem
\bibitem[Giomi(2015)]{Giomi:2014}
L.~Giomi, \emph{Phys. Rev. X}, 2015, \textbf{5}, 031003\relax
\mciteBstWouldAddEndPuncttrue
\mciteSetBstMidEndSepPunct{\mcitedefaultmidpunct}
{\mcitedefaultendpunct}{\mcitedefaultseppunct}\relax
\EndOfBibitem
\bibitem[Thampi \emph{et~al.}(2015)Thampi, Golestanian, and
  Yeomans]{Thampi:2015}
S.~P. Thampi, R.~Golestanian and J.~M. Yeomans, \emph{Molecular Physics}, 2015,
  \textbf{113}, 2656--2665\relax
\mciteBstWouldAddEndPuncttrue
\mciteSetBstMidEndSepPunct{\mcitedefaultmidpunct}
{\mcitedefaultendpunct}{\mcitedefaultseppunct}\relax
\EndOfBibitem
\bibitem[Sanchez \emph{et~al.}(2012)Sanchez, Chen, DeCamp, Heymann, and
  Dogic]{Sanchez:2012}
T.~Sanchez, D.~T.~N. Chen, S.~J. DeCamp, M.~Heymann and Z.~Dogic,
  \emph{Nature}, 2012, \textbf{491}, 431--434\relax
\mciteBstWouldAddEndPuncttrue
\mciteSetBstMidEndSepPunct{\mcitedefaultmidpunct}
{\mcitedefaultendpunct}{\mcitedefaultseppunct}\relax
\EndOfBibitem
\bibitem[Keber \emph{et~al.}(2014)Keber, Loiseau, Sanchez, DeCamp, Giomi,
  Bowick, Marchetti, Dogic, and Bausch]{Keber:2014}
F.~C. Keber, E.~Loiseau, T.~Sanchez, S.~J. DeCamp, L.~Giomi, M.~J. Bowick,
  M.~C. Marchetti, Z.~Dogic and A.~R. Bausch, \emph{Science}, 2014,
  \textbf{345}, 1135\relax
\mciteBstWouldAddEndPuncttrue
\mciteSetBstMidEndSepPunct{\mcitedefaultmidpunct}
{\mcitedefaultendpunct}{\mcitedefaultseppunct}\relax
\EndOfBibitem
\bibitem[Henkin \emph{et~al.}(2014)Henkin, DeCamp, Chen, Sanchez, and
  Dogic]{Henkin:2014}
G.~Henkin, S.~J. DeCamp, D.~T.~N. Chen, T.~Sanchez and Z.~Dogic,
  \emph{Philosophical Transactions of the Royal Society of London A:
  Mathematical, Physical and Engineering Sciences}, 2014, \textbf{372},
  0142\relax
\mciteBstWouldAddEndPuncttrue
\mciteSetBstMidEndSepPunct{\mcitedefaultmidpunct}
{\mcitedefaultendpunct}{\mcitedefaultseppunct}\relax
\EndOfBibitem
\bibitem[Zhou \emph{et~al.}(2013)Zhou, Sokolov, Lavrentovich, and
  Aranson]{Zhou:2014}
S.~Zhou, A.~Sokolov, O.~D. Lavrentovich and I.~S. Aranson, \emph{Proc. Nat.
  Acad. Sci. U.S.A.}, 2013, \textbf{111}, 1265--1270\relax
\mciteBstWouldAddEndPuncttrue
\mciteSetBstMidEndSepPunct{\mcitedefaultmidpunct}
{\mcitedefaultendpunct}{\mcitedefaultseppunct}\relax
\EndOfBibitem
\bibitem[Duclos \emph{et~al.}(2014)Duclos, Garcia, HG, and
  Silberzan]{Duclos:2014}
G.~Duclos, S.~Garcia, Y.~HG and P.~Silberzan, \emph{Soft Matter}, 2014,
  \textbf{10}, 2346--2353\relax
\mciteBstWouldAddEndPuncttrue
\mciteSetBstMidEndSepPunct{\mcitedefaultmidpunct}
{\mcitedefaultendpunct}{\mcitedefaultseppunct}\relax
\EndOfBibitem
\bibitem[Gao \emph{et~al.}(2015)Gao, Blackwell, Glaser, Betterton, and
  Shelley]{Gao:2015}
T.~Gao, R.~Blackwell, M.~A. Glaser, M.~D. Betterton and M.~J. Shelley,
  \emph{Phys. Rev. Lett.}, 2015, \textbf{114}, 048101\relax
\mciteBstWouldAddEndPuncttrue
\mciteSetBstMidEndSepPunct{\mcitedefaultmidpunct}
{\mcitedefaultendpunct}{\mcitedefaultseppunct}\relax
\EndOfBibitem
\bibitem[de~Gennes and Prost(1993)]{DeGennes:1993}
P.~de~Gennes and J.~Prost, \emph{{The Physics of Liquid Crystals}}, Oxford:
  Oxford University Press., 2nd edn., 1993\relax
\mciteBstWouldAddEndPuncttrue
\mciteSetBstMidEndSepPunct{\mcitedefaultmidpunct}
{\mcitedefaultendpunct}{\mcitedefaultseppunct}\relax
\EndOfBibitem
\bibitem[Sokolov \emph{et~al.}(2007)Sokolov, Aranson, Kessler, and
  Goldstein]{Sokolov2007}
A.~Sokolov, I.~S. Aranson, J.~O. Kessler and R.~E. Goldstein, \emph{Phys. Rev.
  Lett.}, 2007, \textbf{98}, 1--4\relax
\mciteBstWouldAddEndPuncttrue
\mciteSetBstMidEndSepPunct{\mcitedefaultmidpunct}
{\mcitedefaultendpunct}{\mcitedefaultseppunct}\relax
\EndOfBibitem
\bibitem[Wensink \emph{et~al.}(2012)Wensink, Dunkel, Heidenreich, Drescher,
  Goldstein, Lowen, and Yeomans]{Wensink2012}
H.~H. Wensink, J.~Dunkel, S.~Heidenreich, K.~Drescher, R.~E. Goldstein,
  H.~Lowen and J.~M. Yeomans, \emph{Proc. Natl. Acad. Sci.}, 2012,
  \textbf{109}, 14308--14313\relax
\mciteBstWouldAddEndPuncttrue
\mciteSetBstMidEndSepPunct{\mcitedefaultmidpunct}
{\mcitedefaultendpunct}{\mcitedefaultseppunct}\relax
\EndOfBibitem
\bibitem[Ravnik and Yeomans(2013)]{Ravnik2013}
M.~Ravnik and J.~M. Yeomans, \emph{Phys. Rev. Lett.}, 2013, \textbf{110},
  026001\relax
\mciteBstWouldAddEndPuncttrue
\mciteSetBstMidEndSepPunct{\mcitedefaultmidpunct}
{\mcitedefaultendpunct}{\mcitedefaultseppunct}\relax
\EndOfBibitem
\bibitem[Olmsted(2008)]{Olmsted:2008}
P.~Olmsted, \emph{Rheologica Acta}, 2008, \textbf{47}, 283--300\relax
\mciteBstWouldAddEndPuncttrue
\mciteSetBstMidEndSepPunct{\mcitedefaultmidpunct}
{\mcitedefaultendpunct}{\mcitedefaultseppunct}\relax
\EndOfBibitem
\bibitem[Hemingway \emph{et~al.}(2015)Hemingway, Maitra, Banerjee, Marchetti,
  Ramaswamy, Fielding, and Cates]{Hemingway:2015}
E.~J. Hemingway, A.~Maitra, S.~Banerjee, M.~C. Marchetti, S.~Ramaswamy, S.~M.
  Fielding and M.~E. Cates, \emph{Physical Review Letters}, 2015, \textbf{114},
  098302\relax
\mciteBstWouldAddEndPuncttrue
\mciteSetBstMidEndSepPunct{\mcitedefaultmidpunct}
{\mcitedefaultendpunct}{\mcitedefaultseppunct}\relax
\EndOfBibitem
\bibitem[Huterer and Vachaspati(2005)]{Huterer:2005}
D.~Huterer and T.~Vachaspati, \emph{Phys.Rev. D}, 2005, \textbf{72},
  043004\relax
\mciteBstWouldAddEndPuncttrue
\mciteSetBstMidEndSepPunct{\mcitedefaultmidpunct}
{\mcitedefaultendpunct}{\mcitedefaultseppunct}\relax
\EndOfBibitem
\bibitem[Fielding \emph{et~al.}(2011)Fielding, Marenduzzo, and
  Cates]{Fielding:2011}
S.~M. Fielding, D.~Marenduzzo and M.~E. Cates, \emph{Physical Review E}, 2011,
  \textbf{83}, 041910\relax
\mciteBstWouldAddEndPuncttrue
\mciteSetBstMidEndSepPunct{\mcitedefaultmidpunct}
{\mcitedefaultendpunct}{\mcitedefaultseppunct}\relax
\EndOfBibitem
\bibitem[Rafa{\"{\i}} \emph{et~al.}(2010)Rafa{\"{\i}}, Jibuti, and
  Peyla]{Rafai2010}
S.~Rafa{\"{\i}}, L.~Jibuti and P.~Peyla, \emph{Phys. Rev. Lett.}, 2010,
  \textbf{104}, 1--4\relax
\mciteBstWouldAddEndPuncttrue
\mciteSetBstMidEndSepPunct{\mcitedefaultmidpunct}
{\mcitedefaultendpunct}{\mcitedefaultseppunct}\relax
\EndOfBibitem
\bibitem[Bendix \emph{et~al.}(2008)Bendix, Koenderink, Cuvelier, Dogic,
  Koeleman, Brieher, Field, Mahadevan, and Weitz]{Bendix2008}
P.~M. Bendix, G.~H. Koenderink, D.~Cuvelier, Z.~Dogic, B.~N. Koeleman, W.~M.
  Brieher, C.~M. Field, L.~Mahadevan and D.~A. Weitz, \emph{Biophys. J.}, 2008,
  \textbf{94}, 3126--3136\relax
\mciteBstWouldAddEndPuncttrue
\mciteSetBstMidEndSepPunct{\mcitedefaultmidpunct}
{\mcitedefaultendpunct}{\mcitedefaultseppunct}\relax
\EndOfBibitem
\bibitem[Dunkel \emph{et~al.}(2013)Dunkel, Heidenreich, Drescher, Wensink,
  B{\"a}r, and Goldstein]{Dunkel2013a}
J.~Dunkel, S.~Heidenreich, K.~Drescher, H.~H. Wensink, M.~B{\"a}r and R.~E.
  Goldstein, \emph{Phys. Rev. Lett.}, 2013, \textbf{110}, 228102\relax
\mciteBstWouldAddEndPuncttrue
\mciteSetBstMidEndSepPunct{\mcitedefaultmidpunct}
{\mcitedefaultendpunct}{\mcitedefaultseppunct}\relax
\EndOfBibitem
\bibitem[Edwards and Yeomans(2009)]{Edwards:2009}
S.~A. Edwards and J.~M. Yeomans, \emph{EPL (Europhysics Letters)}, 2009,
  \textbf{85}, 18008\relax
\mciteBstWouldAddEndPuncttrue
\mciteSetBstMidEndSepPunct{\mcitedefaultmidpunct}
{\mcitedefaultendpunct}{\mcitedefaultseppunct}\relax
\EndOfBibitem
\bibitem[Stark and Lubensky(2003)]{Stark:2003}
H.~Stark and T.~C. Lubensky, \emph{Phys. Rev. E}, 2003, \textbf{67},
  061709\relax
\mciteBstWouldAddEndPuncttrue
\mciteSetBstMidEndSepPunct{\mcitedefaultmidpunct}
{\mcitedefaultendpunct}{\mcitedefaultseppunct}\relax
\EndOfBibitem
\end{mcitethebibliography}
\bibliographystyle{rsc}

\end{document}